\documentclass[twocolumn,aps,prc,superscriptaddress,showpacs,floatfix]{revtex4}

\usepackage{epsfig}
\usepackage{amsmath}
\usepackage{multirow}
\usepackage{graphicx}
\usepackage{amsmath}
\usepackage{feynmf}
\usepackage[normalem]{ulem}  
\usepackage[dvips]{color} 

\renewcommand\sout{\bgroup \color{red} \ULdepth=-.5ex \ULset}

\newcommand{\Ex}[2]{\ifmmode{#1\times10^{#2}}\else{$#1\times10^{#2}$}\fi}

\begin{document}
\title{ 
Dibaryons in a constituent quark model
}

\author{Woosung Park}\affiliation{Department of Physics and Institute of Physics and Applied Physics, Yonsei University, Seoul 120-749, Korea}
\author{Aaron Park}\affiliation{Department of Physics and Institute of Physics and Applied Physics, Yonsei University, Seoul 120-749, Korea}
\author{Su Houng Lee}\affiliation{Department of Physics and Institute of Physics and Applied Physics, Yonsei University, Seoul 120-749, Korea}
\date{\today}
\begin{abstract}
We investigate the properties of dibaryons containing  u and d quarks in the constituent quark model. In constructing  the ground state wave function, we   choose the spatial part to be fully symmetric and the remaining color, isospin and spin part to be antisymmetric so as to satisfy the the Pauli principle.
 By adapting the IS coupling scheme that combine the isospin basis function with the spin basis function, and subsequently coupling this to the color singlet basis function, we construct the color $\otimes$ isospin $\otimes$ spin states compatible with the physical states of the dibaryon.  By using the variational method, we then calculate the mass of the dibaryon in a nonrelativistic potential model, involving Coulomb, color confinement and color-spin hyperfine interaction. In particular, to asses the stability  for  different types of the confinement potential, we introduce one that is linearly proportional to the interquark distance and another to its square root. For all cases considered, we find that there are no compact bound states against the strong decay.
\end{abstract}


\maketitle

\section{Introduction}
The recent obervation of $d^{*}$(2380) with quantum number $I(J^P)$=$0(3^+)$ mesasured by the WASA detector at COSY~\cite{Bashkanov:2008ih,Adlarson:2011bh,Adlarson:2012fe,Adlarson:2014pxj,Adlarson:2014ozl,Adlarson:2014tcn} revived interests in the study of multiquark hadrons, and led to the renwed  investigation of the possible existence of either $\Delta\Delta$ or six-quark state. Theoretically, starting from the work of Jaffe~\cite{Jaffe:1976yi}, there were already many studies on the stability of a six-quark system.  In particular, in relation to $d^{*}$(2380), a study of nonstrange diybaryon was made in Ref.~\cite{Goldman:1989zj}, and a work in Ref.~\cite{Oka:1980ax} was based on one gluon-exchange interaction. Initially, using the Bag  model with strange quark,  Jaffe~\cite{Jaffe:1976yi} predicted that the H-dibaryon with $J^P$=$0^+$ and I=0 consisiting of $uuddss$ could be stable against decay into two  $\Lambda$ baryons, when only the color-spin hyperfine interaction was taken into account.    Using a similar quark model, Silvetre-Brac and Leandri~\cite{SilvestreBrac:1992yg} classified all dibaryon states  within the $SU(3)_F$ representation, and investigated the stability of these states against the decay into the allowed two baryon decay: Through this study, they found  that the $\Omega \Omega$ dibaryon is most likely bound, and H-dibaryon could be stable. 

Additional  models were used to study the stability of the  H-dibaryon; these include lattice gauge~\cite{Mackenzie:1985vv}, bag model~\cite{Aerts:1984vv}, Skyrme model~\cite{Balachandran:1983dj}, and potential model~\cite{Straub:1988mz}.
In Ref.~\cite{Stancu:1998ca}, using the Goldstone boson exchange interaction, the authors predicted that H-dibaryon could not exist.  
While experimental searches for H dibaryon seems to suggest that it is not stable against strong decay for realistic quark masses, recent lattice gauge theory calculations suggest that it does become bound when the quark mass increases\cite{Inoue:2010es,Beane:2010hg}.  

In addition to the study of H-dibaryon, the dibaryon with strangeness -1 or -3 has been proposed by Maltman~\cite{Maltman:1986xz} and Goldman~\cite{Goldman:1987ma}, respectively. Pepin and Stancu~\cite{Pepin:1998ih} investigated the stability of  $uuddsQ$ (Q=$c$ or $b$ ) type of quark configuration  with a chiral constituent quark model, and found the dibaryon to be unstable against strong decay. The  stability of a multiquark system is known to increase when heavier quarks are included in the  tetraquark configuration  ($qq\bar{b}\bar{b}$)~\cite{Zouzou:1986qh,SilvestreBrac:1993ry,Brink:1998as,Park:2013fda}.  Although the mechanism for stability is different, dibaryons with heavy quarks, such as $q^4Q^2$ (Q=$c$ or $b$), have been studied within the simple chromomagnetic model~\cite{Leandri:1995zm}. 

In this paper, we investigate all dibaryon states  containing the u and d quarks, and calculate their  masses in the frame work of nonrelativistic quark model by using the variational method, with a potential that includes the color spin hyperfine potential introduced in Ref.~\cite{Bhaduri:1981pn}.   In order to examine the stability of the dibaryon for strong decay, we first  fit the parameters of the model to reproduce the masses of the baryon multiplet.   Then, by comparing the dibaryon masses to the relevant  two baryons threshold,  we determine whether the  dibaryon is bound  against strong decay. 

The confinement part in our model originates from the effect of one gluon exchange interaction ${\lambda}^c_i{\lambda}^c_j$.  But in principle the Hamiltonian in the SU(3) symmetric quark model can also have a term proportional to the  SU(3) invariant operators in the cubic form which could originate from an intrinsic three-body color confinement interaction. There are two independent three-body color invariant operators; one that can be expressed in terms of two different types of Casimir operator of SU(3), the other that can not. 
 Since we expect that adding a three-body color invariant operators  is very important for the stability of a dibaryon with different flavors, it is necessary to introduce a formula for  the operators in terms of the element of the permutation group $S_6$, based on the  established-formula by Stancu~\cite{Pepin:2001is} and Dmitrasinovic~\cite{Dmitrasinovic:2001nu}.  Using these formula, we calculate the matrix element of the operators with respect to the color singlet basis function in six quarks system, and explore the role of the operators to baryon and dibaryon masses.

This paper is organized as followings. We introduce the Hamiltonian and fit the baryon spectrum in section II. We construct the spatial function in a simple Gaussian form in
section III. We present all of the physical states and construct the color $\otimes$ flavour $\otimes$ states of dibaryon  in section IV. We show the numerical results obtained from the variational method, and deal with the three-body color operators in section V. Finally, we give the summary in Section VI.

\section{Hamiltonian}

For the nonrelativistic Hamiltonian, we take the confinement and hyperfine potential for the color and spin degree of
freedom given by
\begin{eqnarray}
H=\sum_{i=1}^{4}(m_{i}+\frac{\textbf{p}^2_i}{2m_i})-\frac{3}{16}\sum_{i<j}^{4}
\lambda^c_i\lambda^c_j(V^{C}_{ij}+V^{SS}_{ij}), \label{Hamiltonian}
\end{eqnarray}
where $m_i$'s are the quark masses, and $\lambda^c_i/2$ are the color operator of the $i$'th quark for the color SU(3), and $V^{C}_{ij}$ and $V^{SS}_{ij}$ are the confinement and hyperfine potential, respectively. For the  confinement potential, we adopt the following two different types :  
\begin{itemize}
\item Type 1 ;
\begin{eqnarray}
V^{C}_{ij}=-\frac{\kappa}{r_{ij}}+\frac{r_{ij}}{a_0}-D. \label{vc_ij-01}
\end{eqnarray}
In the following analysis, we take the units for  $\kappa$, $a_0$, and D to be $\rm{MeVfm}$, $\rm{(MeV)^{-1}fm}$, and $\rm{MeV}$, respectively.
\item Type 2 ;
\begin{eqnarray}
V^{C}_{ij}=-\frac{\kappa}{r_{ij}}+\frac{(r_{ij})^{1/2}}{a_0}-D. \label{vc_ij-02}
\end{eqnarray}
Here, the units of $a_0$ is taken to be $\rm{(MeV)^{-1}(fm)^{1/2}}$.
\end{itemize}
The hyperfine term which effectively splits the multiplets of baryon with respect to spin is given by, 
\begin{eqnarray}
V^{SS}_{ij}=\frac{\hbar^2c^2{\kappa}^{\prime}}{m_im_jc^4}\frac{1}{(r_{0ij})^2r_{ij}}e^{-(r_{ij})^2/(r_{0ij})^2}
{\sigma}_i\cdot{\sigma}_j, 
\end{eqnarray}
where the unit of ${\kappa}^{\prime}$ is taken to be $\rm{MeVfm}$.
Here, $r_{ij}$ is the distance between interquarks, $\mid\textbf{r}_i-\textbf{r}_j\mid$, and $(r_{0ij})$ are chosen to depend on
the masses of interquarks, given by,
\begin{eqnarray}
r_{0ij}=1/(\alpha+\beta \frac{m_im_j}{m_i+m_j}),
\end{eqnarray}
where the unit of $\alpha$ is $\rm{(fm)^{-1}}$, and the unit of $\beta$, $\rm{(MeVfm)^{-1}}$. We choose to keep the isospin symmetry by requiring that $m_u$=$m_d$ ($\rm{MeV}$). In the Hamiltonian, the parameters have been chosen so that the fitted mass of both baryon octet and decuplet are comparable with those of experiments. 
\begin{table}[htdp]
\caption{ Parameters fitted to the experimental  baryon octet and
decuplet masses for the two different types of potentials.}
\begin{center}
\begin{tabular}{c|c|c|c|c|c|c|c|c|c}
\hline \hline
        & $\kappa$  & $a_0$  & D & ${\kappa}^{\prime}$ & $\alpha$ &$\beta$ & $m_u$  & $m_s$     \\
\hline
Type 1  &  107.6  & 0.001062  & 952.6   & 107.6     &  2.36    & 0.0015        & 340  & 610  \\
\hline
Type 2  &  109.6    & 0.001103  & 963.6   & 168.6     &  2.16   & 0.0018        & 348  & 612  \\
\hline \hline 
\end{tabular}
\end{center}
\label{normalmeson_mass-01}
\end{table}

\subsection{Baryon Spectrum}

In constructing the basis function of a baryon, we restrict  the flavor symmetry to isospin part only, so that we consider only the u, d quarks as identical quarks, and find the total wave function of baryon, according to Pauli principle. When
we calculate the expectation value of the potential terms for baryon with certain symmetry,
it is convenient to introduce the following three Jacobian coordinates so as to reduce our problem to the two body-system in the center
of mass frame :
\begin{itemize}
\item Coordinate I ;
\begin{eqnarray}
\pmb{x_1}=\frac{1}{\sqrt{2}}(\textbf{r}_1-\textbf{r}_2),\qquad
\pmb{x_2}=\sqrt{\frac{2}{3}}(\textbf{r}_3-\frac{1}{2}\textbf{r}_1-\frac{1}{2}\textbf{r}_2).
\end{eqnarray}
\item Coordinate II ;
\begin{eqnarray}
\pmb{y_1}=\frac{1}{\sqrt{2}}(\textbf{r}_1-\textbf{r}_3),\qquad
\pmb{y_2}=\sqrt{\frac{2}{3}}(\textbf{r}_2-\frac{1}{2}\textbf{r}_1-\frac{1}{2}\textbf{r}_3).
\end{eqnarray}
\item Coordinate III ;
\begin{eqnarray}
\pmb{z_1}=\frac{1}{\sqrt{2}}(\textbf{r}_2-\textbf{r}_3),\qquad
\pmb{z_2}=\sqrt{\frac{2}{3}}(\textbf{r}_1-\frac{1}{2}\textbf{r}_2-\frac{1}{2}\textbf{r}_3).
\end{eqnarray}
\end{itemize}

By using simple Gaussian function, we construct the following fully symmetric spatial function  for baryons composed of u and d constituent only :
\begin{align}
R=&\exp[-a(\pmb{x_1})^2-b(\pmb{x_2})^2]+\exp[-a(\pmb{y_1})^2-b(\pmb{y_2})^2]+\nonumber\\
&\exp[-a(\pmb{z_1})^2-b(\pmb{z_2})^2],
\end{align}
where $a$ and $b$ are variational parameters. Since the total wave function of baryon, such as N and P (I=1/2, S=1/2) or $\Delta$ (I=3/2, S=3/2) is fully antisymmetric due to the Pauli principle, the rest of the total wave function must be fully antisymmetric,
if we choose the spatial function to be fully symmetric.  

Concerning the color basis function of the baryon, we consider the color singlet state as  the hadrons  are observed to be colorless. The baryon has only one color singlet state, coming from the irreducible representation of $[3]_C$ $\otimes$ $[3]_C$ $\otimes$ $[3]_C$, given by,
\begin{equation}
\begin{tabular}{c}
$\vert C \rangle$=
\end{tabular}
\begin{tabular}{|c|}
\hline
1   \\
\hline
2 \\
\hline
3  \\
\hline
\end{tabular} 
=\frac{1}{\sqrt{6}}\epsilon_{ijk}q^i(1)q^j(2)q^k(3)),
\end{equation}
which is fully antisymmetric under the exchange of any two particles among 1, 2, and 3. We note
that the Young tableau follows the rule of the standard Young-Yamanouchi representation, which will be shown later in detail.

For the spin basis functions, the baryon can have S=1/2 consisting of two different types, and S=3/2 containing one type as follows :
\begin{itemize}
\item S=3/2 ;
\\
\begin{tabular}{c}
$\vert S^{3/2} \rangle$=
\end{tabular}
\begin{tabular}{|c|c|c|}
\hline
1 & 2 &3  \\
\hline
\end{tabular} 
=$\uparrow\uparrow\uparrow$.
\item S=1/2 ;
\\
\begin{tabular}{c}
$\vert S^{1/2}_1 \rangle$=
\end{tabular}
\begin{tabular}{|c|c|}
\hline
1 & 2   \\
\cline{1-2}
\multicolumn{1}{|c|}{3}  \\
\cline{1-1}
\end{tabular} 
=$\frac{1}{\sqrt{6}}(2\uparrow\uparrow\downarrow-\uparrow\downarrow\uparrow-\downarrow\uparrow\uparrow)$,
\\
\begin{tabular}{c}
$\vert S^{1/2}_2 \rangle$=
\end{tabular}
\begin{tabular}{|c|c|}
\hline
1 & 3  \\
\cline{1-2}
\multicolumn{1}{|c|}{2}  \\
\cline{1-1}
\end{tabular} 
=$\frac{1}{\sqrt{2}}(\uparrow\downarrow\uparrow-\downarrow\uparrow\uparrow)$.
\end{itemize}
As we can see, the spin part of the basis functions of baryon $\vert S^{3/2} \rangle$ is completely symmetric, while that of  $\vert S^{1/2}_1 \rangle$ and $\vert S^{1/2}_2 \rangle$ are partially symmetric; the  former being  symmetric between particle 1 and 2, and 
the latter antisymmetric between particle 1 and 2.

Likewise, we construct the isospin basis function of baryon for I=3/2 and I=1/2 :
\begin{itemize}
\item I=3/2 ;
\\
\begin{tabular}{c}
$\vert I^{3/2} \rangle$=
\end{tabular}
\begin{tabular}{|c|c|c|}
\hline
1 & 2 &3  \\
\hline
\end{tabular} 
=$uuu$.
\item I=1/2 ;
\\
\begin{tabular}{c}
$\vert I^{1/2}_1 \rangle$=
\end{tabular}
\begin{tabular}{|c|c|}
\hline
1 & 2   \\
\cline{1-2}
\multicolumn{1}{|c|}{3}  \\
\cline{1-1}
\end{tabular} 
=$\frac{1}{\sqrt{6}}(2uud-udu-duu)$,
\\
\begin{tabular}{c}
$\vert I^{1/2}_2 \rangle$=
\end{tabular}
\begin{tabular}{|c|c|}
\hline
1 & 3  \\
\cline{1-2}
\multicolumn{1}{|c|}{2}  \\
\cline{1-1}
\end{tabular} 
=$\frac{1}{\sqrt{2}}(udu-duu)$.
\end{itemize}

For the baryons with (I=3/2,S=3/2) and (I=1/2,S=1/2), the antisymmetry property of the total wave function  can be easily obtained from the direct product of the totally antisymmetric part of the color basis function times the totally symmetric properties of the remain function comprised of the spatial function, isospin basis functions, and spin basis functions.  
The fully symmetric part of the isospin $\otimes$ spin basis function in both cases of the isospin and spin and can be written as 
\begin{itemize}
\item $\Delta$ (I=3/2,S=3/2) ;
\\
$\vert I^{3/2}, S^{3/2} \rangle$=$\vert I^{3/2}\rangle \otimes \vert S^{3/2}\rangle$,
\item N, P (I=1/2,S=1/2) ;
\\
$\vert I^{1/2}, S^{1/2} \rangle$=$\frac{1}{\sqrt{2}}(\vert I^{1/2}_1\rangle \otimes \vert S^{1/2}_1\rangle+\vert I^{1/2}_2\rangle \otimes \vert S^{1/2}_2\rangle)$. 
\end{itemize}
In the case of (I=1/2,S=1/2), we can see that $\vert I^{1/2}, S^{1/2} \rangle$ is symmetric between particle 1 and 2. The remaining symmetry for the permutation (23) can be deduced from the following formulas, according to the rule to the standard Young-Yamanouchi representation :

\begin{equation}
\begin{tabular}{c}
(23)
\end{tabular}
\begin{tabular}{|c|c|}
\hline
1 & 2   \\
\cline{1-2}
\multicolumn{1}{|c|}{3}  \\
\cline{1-1}
\end{tabular} 
=-\frac{1}{2}
\begin{tabular}{|c|c|}
\hline
1 & 2   \\
\cline{1-2}
\multicolumn{1}{|c|}{3}  \\
\cline{1-1}
\end{tabular}
+\frac{\sqrt{3}}{2}
\begin{tabular}{|c|c|}
\hline
1 & 3  \\
\cline{1-2}
\multicolumn{1}{|c|}{2}  \\
\cline{1-1}
\end{tabular}\nonumber
\end{equation} 
\begin{equation}
\begin{tabular}{c}
(23)
\end{tabular}
\begin{tabular}{|c|c|}
\hline
1 & 3   \\
\cline{1-2}
\multicolumn{1}{|c|}{2}  \\
\cline{1-1}
\end{tabular} 
=\frac{1}{2}
\begin{tabular}{|c|c|}
\hline
1 & 3   \\
\cline{1-2}
\multicolumn{1}{|c|}{2}  \\
\cline{1-1}
\end{tabular}
+\frac{\sqrt{3}}{2}
\begin{tabular}{|c|c|}
\hline
1 & 2  \\
\cline{1-2}
\multicolumn{1}{|c|}{3}  \\
\cline{1-1}
\end{tabular} 
\end{equation}

Now, we can construct the basis function of the color $\otimes$ isospin $\otimes$ spin for (I=3/2,S=3/2) and (I=1/2,S=1/2) that are completely antisymmetric.  These are given by 
\begin{align}
\vert I^{3/2}, S^{3/2} \rangle=\vert C \rangle \otimes \vert I^{3/2}, S^{3/2} \rangle, \nonumber \\
\vert I^{1/2}, S^{1/2} \rangle=\vert C \rangle \otimes \vert I^{1/2}, S^{1/2} \rangle.
\end{align}

For the hyperon, we treat the strange quark as distinguishable from the u and d quarks and will  
not require  the total wave function to be fully antisymmetric.    It is then easy to construct the total
wave function for the hyperons with strangeness s=-1 and s=-2, that satisfy the  Pauli principle in the u and d quark sectors only.
\begin{table}[htdp]
\caption{This table shows the mass of baryons in octet and decuplet obtained from the variational method. The fourth row indicates the experimental data. ( unit ; $\rm{MeV}$ ) }
\begin{center}
\begin{tabular}{c|c|c|c|c|c|c|c}
\hline \hline
(I,S)      & ($\frac{1}{2}$,$\frac{1}{2}$) &($\frac{1}{2}$,$\frac{1}{2}$)& (0,$\frac{1}{2}$) & (1,$\frac{1}{2}$) &($\frac{1}{2}$,$\frac{3}{2}$) &(1,$\frac{3}{2}$) &($\frac{3}{2}$,$\frac{3}{2}$)   \\
           & N, P     & $\Xi$    & $\Lambda$   & $\Sigma$     & $\Xi^*$   & $\Sigma^*$    &  $\Delta$                 \\
\hline
Type 1  &  977.1   & 1315.3   & 1115.6        & 1206.0     & 1530.2       & 1403.4    & 1267.5 \\                     
\hline
Type 2  &  976.3    & 1380.3  & 1115.6   & 1238.2   &  1593.0   & 1419.3        & 1237.2 \\
\hline
Exp       &  938.2   &  1314.8    &              & 1189.3    & 1530     & 1382.8       & 1230        \\
            & $\sim$  & $\sim$     & 1115.6  & $\sim$    &  $\sim$ & $\sim$    & $\sim$ \\
          &  939.5   &  1321.7        &              & 1197.4     & 1531.8  & 1387.2      & 1234       \\  
\hline \hline
\end{tabular}
\end{center}
\label{normalmeson_mass-02}
\end{table}

\section{ Spatial function}

In order to construct a fully antisymmetric wave function of dibaryon containing only identical u and d  
particles, we choose the spatial function to be fully symmetric such that the  rest of the wave function represented by color $\otimes$ flavor $\otimes$ spin should be antisymmetric.
Since we restrict the SU(3)  flavor of dibaryon to the  isospin symmetry only, the flavor state can be identified with the isospin quantum number.   
In describing the system consisting of six quarks, it is convenient to deal with the system in the center of mass frame, reducing the number of suitable Jacobian coordinates of the system to five. The five Jacobian coordinates are given by
\begin{align}
&\pmb{x_1^1}=\frac{1}{\sqrt{2}}(\textbf{r}_1-\textbf{r}_2),\quad
\pmb{x_2^1}=\frac{1}{2}(\textbf{r}_3-\textbf{r}_4+\textbf{r}_5-\textbf{r}_6),\nonumber\\
&\pmb{x_3^1}=\frac{1}{2}(\textbf{r}_3-\textbf{r}_4-\textbf{r}_5+\textbf{r}_6),\quad
\pmb{x_4^1}=\frac{1}{2}(\textbf{r}_3+\textbf{r}_4-\textbf{r}_5-\textbf{r}_6),\nonumber\\
&\pmb{x_5^1}=\frac{1}{\sqrt{12}}(\textbf{r}_3+\textbf{r}_4+\textbf{r}_5+\textbf{r}_6-2\textbf{r}_1-2\textbf{r}_2).
\label{eq-jac1}
\end{align}
The variational method for calculating the mass of dibaryon turns out to be easy when the  Gaussian form with respect to the Jacobian coordinates are used for the spatial wave function.  
Using the Jacobian coordinates given in Eq.~(\ref{eq-jac1}) in the Gaussian wave function, we find the form to be  symmetric under the exchange of any two particles among 3, 4, 5, 6,  and at the same time symmetric under the exchange of two particles between 1 and 2; these symmetry properties are denoted as [3456][12]. Introducing the variational parameters $a$, $b$, $c$,  the spatial function is then given by  
\begin{align}
R^{s_1}=\exp[-&(a(\pmb{x_1^1})^2+b(\pmb{x_2^1})^2+b(\pmb{x_3^1})^2+b(\pmb{x_4^1})^2+\nonumber\\
&c(\pmb{x_5^1})^2)].
\end{align}

In addition to this Gaussian function, the full symmetry of the spatial function requires the linear sum of 14 additional Gaussian functions, each of which has a specific symmetry under particle exchange. The next set of five Jacobian coordinates with the
symmetry of [2456][13] is given by  
\begin{align}
&\pmb{x_1^2}=\frac{1}{\sqrt{2}}(\textbf{r}_1-\textbf{r}_3),\quad
\pmb{x_2^2}=\frac{1}{2}(\textbf{r}_2-\textbf{r}_4+\textbf{r}_5-\textbf{r}_6),\nonumber\\
&\pmb{x_3^2}=\frac{1}{2}(\textbf{r}_2-\textbf{r}_4-\textbf{r}_5+\textbf{r}_6),\quad
\pmb{x_4^2}=\frac{1}{2}(\textbf{r}_2+\textbf{r}_4-\textbf{r}_5-\textbf{r}_6),\nonumber\\
&\pmb{x_5^2}=\frac{1}{\sqrt{12}}(\textbf{r}_2+\textbf{r}_4+\textbf{r}_5+\textbf{r}_6-2\textbf{r}_1-2\textbf{r}_3).
\end{align}
The corresponding  Gaussian function specifying the symmetry of [2456][13] is given by  
\begin{align}
R^{s_2}=\exp[-&(a(\pmb{x_1^2})^2+b(\pmb{x_2^2})^2+b(\pmb{x_3^2})^2+b(\pmb{x_4^2})^2+\nonumber\\
&c(\pmb{x_5^2})^2)]. \label{space-wave-function2}
\end{align}
We find that the set of Jacobian coordinates necessary to obtain the  fully symmetric wave function under the exchange of any two particles among 1, 2, 3, 4, 5, and 6, and consequently the corresponding Gaussian functions, are the ones with the following symmetry; [3456][12], [2456][13], [2356][14], [2346][15], [2345][16], [1456][23], [1356][24], [1346][25], [1345][26], [1256][34], [1246][35], [1245][36], [1236][45], [1235][46],   [1234][56].

Combining these Gaussian functions with the symmetry into a linear form, we obtain the spatial function with three variational parameters $a$, $b$, $c$ which is fully symmetric as follows  
\begin{align}
R^{s}=&R^{s_1}+R^{s_2}+R^{s_3}+R^{s_4}+R^{s_5}+\nonumber\\
&R^{s_6}+R^{s_7}+R^{s_8}+R^{s_{9}}+R^{s_{10}}+\nonumber\\
&R^{s_{11}}+R^{s_{12}}+R^{s_{13}}+R^{s_{14}}+R^{s_{15}}.
\end{align}
It is easy to check the symmetry of the spatial function with respect to all of the permutation of $S_6$, by considering only the 5 permutations  (12), (23), (34), (45), and (56), as these permutations generates all of the permutations of $S_6$.

\section{Classification of dibaryon with isospin symmetry}

\subsection{Isospin and spin state of dibaryon}
In this section, we investigate the state of the dibaryon consisting of identical  u, d quarks, whose flavor part is characterized by 
isospin symmetry. Since the color $\otimes$ isospin $\otimes$ spin state of each quark  can be represented by $[3]_{C}$ $\otimes$ $[2]_{I}$ $\otimes$ $[2]_{S}$,  the direct product of six quarks enable us to classify all of the states of dibaryon with respect to the state of isospin and spin, denoted by $\vert I, S \rangle$. In our notation, the $[3]_{C}$ indicate the fundamental representation of $SU(3)_{C}$,
$[2]_{I}$ the fundamental representation of $SU(2)_{I}$, and $[2]_{S}$ the fundamental representation of $SU(2)_{S}$. In our case where we choose the spatial function of dibaryon to be fully symmetric, the color $\otimes$ isospin $\otimes$ spin state of the dibaryon will be chosen to be  fully antisymmetric. The fully antisymmetric state of the color $\otimes$ isospin $\otimes$ spin state can be easily obtained from the classifying of the multiplets of the direct six product of $[12]_{CIS}$, which is the fundamental representation of  $SU(12)_{CIS}$, and gives the multiplet with dimension 924 represented by Young tableau $[1^6]$. 
Using the original representation of $[3]_{C}$ $\otimes$ $[4]_{IS}$, which we will equivalently represent as ($[3]_{C}$, $[4]_{IS}$), the totally antisymmetric multiplet of $[1^6]_{CIS}$ can be 
decomposed as
\begin{align}      
[1^6]_{CIS}=&([1]_{C}, [50]_{IS}) \oplus ([8]_{C}, [64]_{IS}) \oplus ([\bar{10}]_{C}, [\bar{10}]_{IS}) \oplus \nonumber\\
&([10]_{C}, [10]_{IS}) \oplus ([27]_{C}, [6]_{IS}). 
\label{eq-1-6}
\end{align}
By using the Young tableau, we can easily find that the multiplets in the right hand side of Eq.~(18) is fully antisymmetric. According to the 
permutation group theory, for a given Young tableau, the fully antisymmetric function can be constructed by multiplying the 
Young tableau by its conjugate of the Young tableau, where the conjugate representation of a given Young 
tableau can be obtained by exchanging the row and column in the Young tableau :
 
\begin{center}
\begin{tabular}{c}
$([1]_{C}, [50]_{IS})$=
\end{tabular}
\begin{tabular}{|c|c|}
\hline
$\quad$ & $\quad$   \\
\cline{1-2}
$\quad$ &  $\quad$  \\
\cline{1-2}
$\quad$ &  $\quad$  \\
\hline
\end{tabular}
$\otimes$
\begin{tabular}{|c|c|c|}
\hline
$\quad$ & $\quad$ & $\quad$    \\
\cline{1-3} $\quad$  & $\quad$ & $\quad$  \\
\hline
\end{tabular}
,
\end{center}

\begin{center}
\begin{tabular}{c}
$([8]_{C}, [64]_{IS})$=
\end{tabular}
\begin{tabular}{|c|c|c|}
\hline
$\quad$ & $\quad$ & $\quad$   \\
\hline
\multicolumn{1}{|c|}{$\quad$} & \multicolumn{1}{c|}{$\quad$} \\
\cline{1-2}
\multicolumn{1}{|c|}{$\quad$}   \\
\cline{1-1}
\end{tabular}
$\otimes$
\begin{tabular}{|c|c|c|}
\hline
$\quad$ & $\quad$ & $\quad$   \\
\hline
\multicolumn{1}{|c|}{$\quad$} & \multicolumn{1}{c|}{$\quad$} \\
\cline{1-2}
\multicolumn{1}{|c|}{$\quad$}   \\
\cline{1-1}
\end{tabular}
,
\end{center}

\begin{center}
\begin{tabular}{c}
$([\bar{10}]_{C}, [\bar{10}]_{IS})$=
\end{tabular}
\begin{tabular}{|c|c|c|}
\hline
$\quad$ & $\quad$ & $\quad$    \\
\cline{1-3} 
$\quad$  & $\quad$ & $\quad$  \\
\hline
\end{tabular}
$\otimes$
\begin{tabular}{|c|c|}
\hline
$\quad$ & $\quad$   \\
\cline{1-2}
$\quad$ &  $\quad$  \\
\cline{1-2}
$\quad$ &  $\quad$  \\
\hline
\end{tabular}
,
\end{center}


\begin{center}
\begin{tabular}{c}
$([10]_{C}, [10]_{IS})$=
\end{tabular}
\begin{tabular}{|c|c|c|c|}
\hline
$\quad$ & $\quad$ & $\quad$ & $\quad$   \\
\cline{1-4}
\multicolumn{1}{|c|}{$\quad$}   \\
\cline{1-1}
\multicolumn{1}{|c|}{$\quad$}   \\
\cline{1-1}
\end{tabular}
$\otimes$
\begin{tabular}{|c|c|c|}
\hline
$\quad$ & $\quad$ & $\quad$   \\
\cline{1-3}
\multicolumn{1}{|c|}{$\quad$}   \\
\cline{1-1}
\multicolumn{1}{|c|}{$\quad$}   \\
\cline{1-1}
\multicolumn{1}{|c|}{$\quad$}   \\
\cline{1-1}
\end{tabular}
,
\end{center}



\begin{center}
\begin{tabular}{c}
$([27]_{C}, [6]_{IS})$=
\end{tabular}
\begin{tabular}{|c|c|c|c|}
\hline
$\quad$ & $\quad$ & $\quad$ & $\quad$   \\
\cline{1-4}
\multicolumn{1}{|c|}{$\quad$} & \multicolumn{1}{c|}{$\quad$}  \\
\cline{1-2}
\end{tabular}
$\otimes$
\begin{tabular}{|c|c|}
\hline
$\quad$ & $\quad$   \\
\hline
$\quad$ & $\quad$   \\
\hline
\multicolumn{1}{|c|}{$\quad$}  \\
\cline{1-1}
\multicolumn{1}{|c|}{$\quad$}  \\
\cline{1-1}
\end{tabular}
.
\end{center}


Because the dibaryon is supposed to be a physically observable color singlet state, the dibaryon belongs to the five independent color singlet states represented by the Young tableau of [2,2,2] from the classification of the multiplets of the direct six times of $[3]_{C}$. Hence, only $([1]_{C}, [50]_{IS})$ states  in Eq.~(\ref{eq-1-6}) is allowed as the physical states of the dibaryon. We
can also find the decomposition of $[50]_{IS}$ with respect to the multiplets of $[2]_I$ $\otimes$ $[2]_S$, given by,
\begin{align}      
[50]_{IS}=&([1]_{I}, [3]_{S}) \oplus ([3]_{I}, [1]_{S}) \oplus ([3]_{I}, [5]_{S}) \oplus \nonumber\\
&([5]_{I}, [3]_{S}) \oplus ([7]_{I}, [1]_{S}) \oplus ([1]_{I}, [7]_{S})    .
\label{eq-50is} 
\end{align}
These $\vert I, S \rangle$ states are all the possible states of dibaryon with isospin symmetry which satisfy the antisymmetry property and the color singlet requirement as an observable. As we see in the Eq.~(\ref{eq-50is}), the isospin and spin
of the dibaryon with isospin symmetry has I=0(S=0), I=1(S=1), I=2(S=2), and I=3(S=3), coming from the classification
of SU(2) of the dibaryon, given by,  
\begin{align}      
&[2]_I\otimes[2]_I\otimes[2]_I\otimes[2]_I\otimes[2]_I\otimes[2]_I=\nonumber\\
&[1]_{I=0} \otimes F_{[3,3]} \oplus [3]_{I=1} \otimes F_{[4,2]} \oplus \nonumber\\
&[5]_{I=2} \otimes F_{[5,1]} \oplus [7]_{I=3} \otimes F_{[6]},
\end{align} 
where F indicates the number of times  the corresponding state (I) appears in the product.  Moreover,  the subscript of F represents the Young tableau for the state (I), $F_{[3,3]}$=5, $F_{[4,2]}$=9, $F_{[5,1]}$=5, and $F_{[6]}$=1.  Consequently, there
are 6 different kind of $\vert I, S \rangle$ states of the dibaryon which 
dictates the possible two baryon decay mode in strong force. 
In order to investigate the stability of the dibaryon against a strong decay, we will examine energies in relation to the threshold for the decay mode.
\begin{table}[htdp]
\caption{The decay mode for the dibaryon into two baryons with respect to (I,S) states.
}
\begin{center}
\begin{tabular}{c|c|c|c|c|c|c}
\hline \hline
        & $\Delta\Delta$  & $\Delta$N  & $\Delta$N & NN & $\Delta\Delta$ & NN  \\
\hline
(I,S)  &  (3,0)               &  (2,1)         & (1,2)        & (1,0)     &  (0,3)        & (0,1)  \\
\hline \hline 
\end{tabular}
\end{center}
\label{normalmeson_mass-03}
\end{table}

It is very important to understand the property of basis functions of the dibaryon in calculating the expectation values
of both the confinement and hyperfine potential, proportional to either $\lambda_i^c\lambda_j^c$ or $\lambda_i^c\lambda_j^c{\sigma}_i\cdot{\sigma}_j$, respectively. So we will now establish the basis functions of the color, isospin, 
and spin based on the Young tableau, which are very useful for constructing completely antisymmetric states. Then,
the expectation values can be easily calculated by using the complete antisymmetry properties.

\subsection{Color basis functions}

The dibaryon of our interest with isospin symmetry has color singlet function represented by the 
Young tableau of [2,2,2], as we mentioned earlier. Since the dimension of the Young tableau of [2,2,2] is five, there are five color singlet functions corresponding to the Young tableau of [2,2,2]. We define the color singlet functions as the followings : 
\begin{align}
&\begin{tabular}{c}
$\vert C_1 \rangle$=
\end{tabular}
\begin{tabular}{|c|c|}
\hline
1 & 2   \\
\cline{1-2}
3 &  4  \\
\cline{1-2}
5 &  6  \\
\hline
\end{tabular} 
=\{[(12)_63]_8[4(56)_6]_8\}_1=\nonumber
\\
&\frac{1}{3\sqrt{2}}(\epsilon_{ijk}q^i(1)q^j(3)q^k(5)\epsilon_{lmn}q^l(2)q^m(4)q^n(6) \nonumber\\
&-\frac{1}{2}\epsilon_{ijk}q^i(1)q^j(2)q^k(5)\epsilon_{lmn}q^l(3)q^m(4)q^n(6) \nonumber\\                                                                                                                     
&-\frac{1}{2}\epsilon_{ijk}q^i(1)q^j(3)q^k(4)\epsilon_{lmn}q^l(2)q^m(5)q^n(6) \nonumber\\                                                                                                                     &+\frac{1}{4}\epsilon_{ijk}q^i(1)q^j(2)q^k(4)\epsilon_{lmn}q^l(3)q^m(5)q^n(6) \nonumber\\                                                                                                                    
&-\frac{3}{4}\epsilon_{ijk}q^i(1)q^j(2)q^k(3)\epsilon_{lmn}q^l(4)q^m(5)q^n(6)), \nonumber                                                                                                                                                                                                                                                                                                                                                    
\end{align}
\begin{align}
&\begin{tabular}{c}
$\vert C_2 \rangle$=
\end{tabular}
\begin{tabular}{|c|c|}
\hline
1 & 3   \\
\cline{1-2}
2 &  4  \\
\cline{1-2}
5 &  6  \\
\hline
\end{tabular} 
=\{[(12)_{\bar{3}}3]_8[4(56)_6]_8\}_1=\nonumber
\\
&\frac{1}{2\sqrt{6}}(\epsilon_{ijk}q^i(1)q^j(2)q^k(5)\epsilon_{lmn}q^l(3)q^m(4)q^n(6) \nonumber\\
&-\frac{1}{2}\epsilon_{ijk}q^i(1)q^j(2)q^k(4)\epsilon_{lmn}q^l(3)q^m(5)q^n(6) \nonumber\\                                                                                                                     
&+\frac{1}{2}\epsilon_{ijk}q^i(1)q^j(2)q^k(3)\epsilon_{lmn}q^l(4)q^m(5)q^n(6)), \nonumber                                                                                                                                                                                                                                 
\end{align}
\begin{align}
&\begin{tabular}{c}
$\vert C_3 \rangle$=
\end{tabular}
\begin{tabular}{|c|c|}
\hline
1 & 2   \\
\cline{1-2}
3 &  5  \\
\cline{1-2}
4 &  6  \\
\hline
\end{tabular} 
=\{[(12)_63]_8[4(56)_{\bar{3}}]_8\}_1=\nonumber
\\
&\frac{1}{2\sqrt{6}}(\epsilon_{ijk}q^i(1)q^j(3)q^k(4)\epsilon_{lmn}q^l(2)q^m(5)q^n(6) \nonumber\\
&-\frac{1}{2}\epsilon_{ijk}q^i(1)q^j(2)q^k(4)\epsilon_{lmn}q^l(3)q^m(5)q^n(6) \nonumber\\                                                                                                                     
&+\frac{1}{2}\epsilon_{ijk}q^i(1)q^j(2)q^k(3)\epsilon_{lmn}q^l(4)q^m(5)q^n(6)), \nonumber                                                                                                                                                                                                                                 
\end{align}
\begin{align}
&\begin{tabular}{c}
$\vert C_4 \rangle$=
\end{tabular}
\begin{tabular}{|c|c|}
\hline
1 & 3   \\
\cline{1-2}
2 &  5  \\
\cline{1-2}
4 &  6  \\
\hline
\end{tabular} 
=\{[(12)_{\bar{3}}3]_8[4(56)_{\bar{3}}]_8\}_1=\nonumber
\\
&\frac{1}{4\sqrt{2}}(\epsilon_{ijk}q^i(1)q^j(2)q^k(4)\epsilon_{lmn}q^l(3)q^m(5)q^n(6) \nonumber\\
&-\frac{1}{3}\epsilon_{ijk}q^i(1)q^j(2)q^k(3)\epsilon_{lmn}q^l(4)q^m(5)q^n(6)), \nonumber                                                                                                                                                                                                                                  
\end{align}
\begin{align}
&\begin{tabular}{c}
$\vert C_5 \rangle$=
\end{tabular}
\begin{tabular}{|c|c|}
\hline
1 & 4   \\
\cline{1-2}
2 &  5  \\
\cline{1-2}
3 &  6  \\
\hline
\end{tabular} 
=\{[(12)_{\bar{3}}3]_1[4(56)_{\bar{3}}]_1\}_1=\nonumber
\\
&\frac{1}{6}\epsilon_{ijk}q^i(1)q^j(2)q^k(3)\epsilon_{lmn}q^l(4)q^m(5)q^n(6).                                                                                                              
\end{align}
We note that the color singlet functions followed by the standard Young-Yamanouchi representation is symmetric
with respect to any adjacent particles that lie in the same row, and is antisymmetric with respect to any particles
that lie in the same column. The definition next to the Young tableau expresses the convenient intermediate states for constructing the color singlet, and originates from the [8]$\otimes$[8], and the [1]$\otimes$[1]. The orthogonality of the color singlet functions, $\langle C_i \vert C_j \rangle$=${\delta}_{ij}$ is easily obtained by using the tensor form, and in fact, results from the orthogonality of the standard Young-Yamanouchi bases.

 There are several  ways for calculating  the expectation value of $\lambda_i^c\lambda_j^c$ with respect to the color singlet functions. Among those, it is very useful to consider one that is based on the irreducible matrix representation
of the permutation group with respect to the standard Young-Yamanouchi bases whose irreducible matrix for the transposition is symmetric. Moreover, when considering the operator for describing three gluon exchange, through either $if_{abc}\lambda_i^a\lambda_j^b\lambda_j^c$, or $d_{abc}\lambda_i^a\lambda_j^b\lambda_j^c$, as we shall show the detailed calculation in Appendix, this method gives us a simple form if we know the irreducible matrix representation
of the permutation group. However, in our case of the fully antisymmetric color $\otimes$ isospin $\otimes$ spin state, labelled by $\vert C_iI_jS_k \rangle$, we only need the expectation value of $\lambda_1^c\lambda_2^c$ because the expectation value of $\lambda_i^c\lambda_j^c$ as to a possible
as the expectation value of $\lambda_i^c\lambda_j^c$ can be obtained from the former using the  antisymmetric property of the  basis functions.
In fact, this calculation
is performed using the formula, $\sum_{i<j}^{6}\lambda_i^c\lambda_j^c$=-8/3N, where N is the number of the 
participant particle in this dibaryon, N=6, resulting in 15$\langle \lambda_1^c\lambda_2^c \rangle$=-16. Here, 15 is
the number of ways of pairing between particle i and particle j ($i < j$, i, j =1, 2, 3, 4, 5, and 6 ).

\subsection{Flavor basis functions}

Since the flavor of the dibaryon is in the irreducible representation of SU(2), we consider the flavor basis functions in terms of the isospin representation that is allowed to the dibaryon. 
As in the case of color basis functions, the isospin part can be obtained using similar techniques based on Young tableau. Moreover, it is convenient to establish the orthogonal basis functions with a certain symmetry, making use of the standard Young-Yamanouchi bases.
For the dibaryon, as mentioned above, the representation of isospin has I=0, whose  Young tableau is [3,3] with dimension of 5, and I=1, whose Young tableau is [4,2] with dimension of 9, and I=2, whose Young tableau is [5,1] with dimension of 5, and I=3, whose Young tableau is [6] with dimension of 1 : 
\begin{itemize}
\item I=0 : 5 basis functions with Young tableau [3,3] 
\begin{tabular}{c}
$\vert I_1^0 \rangle$=
\end{tabular}
\begin{tabular}{|c|c|c|}
\hline
                1   & 2   & 3    \\
\cline{1-3} 4  &  5 & 6  \\
\hline
\end{tabular}
\begin{tabular}{c}
$\vert I_2^0 \rangle$=
\end{tabular}
\begin{tabular}{|c|c|c|}
\hline
                1   & 2   & 4    \\
\cline{1-3} 3  &  5 & 6  \\
\hline
\end{tabular}
\begin{tabular}{c}
$\vert I_3^0 \rangle$=
\end{tabular}
\begin{tabular}{|c|c|c|}
\hline
                1   & 3   & 4    \\
\cline{1-3} 2  &  5 & 6  \\
\hline
\end{tabular}

\begin{tabular}{c}
$\vert I_4^0 \rangle$=
\end{tabular}
\begin{tabular}{|c|c|c|}
\hline
                1   & 2   & 5    \\
\cline{1-3} 3  &  4 & 6  \\
\hline
\end{tabular}
\begin{tabular}{c}
$\vert I_5^0 \rangle$=
\end{tabular}
\begin{tabular}{|c|c|c|}
\hline
                1   & 3   & 5    \\
\cline{1-3} 2  &  4 & 6  \\
\hline
\end{tabular}
\item I=1 : 9 basis functions with Young tableau [4,2]
\begin{tabular}{c}
$\vert I_1^1 \rangle$=
\end{tabular}
\begin{tabular}{|c|c|c|c|}
\hline
                  1 & 2 & 3 & 4   \\
\cline{1-4}
\multicolumn{1}{|c|}{5} & \multicolumn{1}{c|}{6}  \\
\cline{1-2}
\end{tabular}
\begin{tabular}{c}
$\vert I_2^1 \rangle$=
\end{tabular}
\begin{tabular}{|c|c|c|c|}
\hline
                  1 & 2 & 3 & 5   \\
\cline{1-4}
\multicolumn{1}{|c|}{4} & \multicolumn{1}{c|}{6}  \\
\cline{1-2}
\end{tabular}
\begin{tabular}{c}
$\vert I_3^1 \rangle$=
\end{tabular}
\begin{tabular}{|c|c|c|c|}
\hline
                  1 & 2 & 4 & 5   \\
\cline{1-4}
\multicolumn{1}{|c|}{3} & \multicolumn{1}{c|}{6}  \\
\cline{1-2}
\end{tabular}

\begin{tabular}{c}
$\vert I_4^1 \rangle$=
\end{tabular}
\begin{tabular}{|c|c|c|c|}
\hline
                  1 & 3 & 4 & 5   \\
\cline{1-4}
\multicolumn{1}{|c|}{2} & \multicolumn{1}{c|}{6}  \\
\cline{1-2}
\end{tabular}
\begin{tabular}{c}
$\vert I_5^1 \rangle$=
\end{tabular}
\begin{tabular}{|c|c|c|c|}
\hline
                  1 & 2 & 3 & 6   \\
\cline{1-4}
\multicolumn{1}{|c|}{4} & \multicolumn{1}{c|}{5}  \\
\cline{1-2}
\end{tabular}
\begin{tabular}{c}
$\vert I_6^1 \rangle$=
\end{tabular}
\begin{tabular}{|c|c|c|c|}
\hline
                  1 & 2 & 4 & 6   \\
\cline{1-4}
\multicolumn{1}{|c|}{3} & \multicolumn{1}{c|}{5}  \\
\cline{1-2}
\end{tabular}

\begin{tabular}{c}
$\vert I_7^1 \rangle$=
\end{tabular}
\begin{tabular}{|c|c|c|c|}
\hline
                  1 & 3 & 4 & 6   \\
\cline{1-4}
\multicolumn{1}{|c|}{2} & \multicolumn{1}{c|}{5}  \\
\cline{1-2}
\end{tabular}
\begin{tabular}{c}
$\vert I_8^1 \rangle$=
\end{tabular}
\begin{tabular}{|c|c|c|c|}
\hline
                  1 & 2 & 5 & 6   \\
\cline{1-4}
\multicolumn{1}{|c|}{3} & \multicolumn{1}{c|}{4}  \\
\cline{1-2}
\end{tabular}
\begin{tabular}{c}
$\vert I_9^1 \rangle$=
\end{tabular}
\begin{tabular}{|c|c|c|c|}
\hline
                  1 & 3 & 5 & 6   \\
\cline{1-4}
\multicolumn{1}{|c|}{2} & \multicolumn{1}{c|}{4}  \\
\cline{1-2}
\end{tabular}
\item I=2 : 5 basis functions with Young tableau [5,1] 
\begin{tabular}{c}
$\vert I_1^2 \rangle$=
\end{tabular}
\begin{tabular}{|c|c|c|c|c|}
\hline
                  1 & 2 & 3 & 4 & 5   \\
\cline{1-5}
\multicolumn{1}{|c|}{6}  \\
\cline{1-1}
\end{tabular}
\begin{tabular}{c}
$\vert I_2^2 \rangle$=
\end{tabular}
\begin{tabular}{|c|c|c|c|c|}
\hline
                  1 & 2 & 3 & 4 & 6   \\
\cline{1-5}
\multicolumn{1}{|c|}{5}  \\
\cline{1-1}
\end{tabular}

\begin{tabular}{c}
$\vert I_3^2 \rangle$=
\end{tabular}
\begin{tabular}{|c|c|c|c|c|}
\hline
                  1 & 2 & 3 & 5 & 6   \\
\cline{1-5}
\multicolumn{1}{|c|}{4}  \\
\cline{1-1}
\end{tabular}
\begin{tabular}{c}
$\vert I_4^2 \rangle$=
\end{tabular}
\begin{tabular}{|c|c|c|c|c|}
\hline
                  1 & 2 & 4 & 5 & 6   \\
\cline{1-5}
\multicolumn{1}{|c|}{3}  \\
\cline{1-1}
\end{tabular}

\begin{tabular}{c}
$\vert I_5^2 \rangle$=
\end{tabular}
\begin{tabular}{|c|c|c|c|c|}
\hline
                  1 & 3 & 4 & 5 & 6   \\
\cline{1-5}
\multicolumn{1}{|c|}{2}  \\
\cline{1-1}
\end{tabular}
\item I=3 : 1 basis function with Young tableau [6] 
\begin{tabular}{c}
$\vert I^3 \rangle$=
\end{tabular}
\begin{tabular}{|c|c|c|c|c|c|}
\hline
                  1 & 2 & 3 & 4 & 5 & 6   \\
\hline
\end{tabular}
\end{itemize}
We can also see the orthogonality of the isospin states, $\langle I_i^l \vert I_j^l \rangle$ = ${\delta}_{ij}$ in a given irreducible representation of isospin from the orthogonality of the standard Young-Yamanouchi bases as well as the orthogonality between any two different irreducible representations, according to the group theory.

\subsection{Spin basis functions}

For the spin states of the dibaryon, the representation of spin states of dibaryon is the same as that of the isosopin states because the irreducible representation of SU(2) should also be applied in this case :
\begin{itemize}
\item S=0 : 5 basis functions with Young tableau [3,3] 
\begin{tabular}{c}
$\vert S_1^0 \rangle$=
\end{tabular}
\begin{tabular}{|c|c|c|}
\hline
                1   & 2   & 3    \\
\cline{1-3} 4  &  5 & 6  \\
\hline
\end{tabular}
\begin{tabular}{c}
$\vert S_2^0 \rangle$=
\end{tabular}
\begin{tabular}{|c|c|c|}
\hline
                1   & 2   & 4    \\
\cline{1-3} 3  &  5 & 6  \\
\hline
\end{tabular}
\begin{tabular}{c}
$\vert S_3^0 \rangle$=
\end{tabular}
\begin{tabular}{|c|c|c|}
\hline
                1   & 3   & 4    \\
\cline{1-3} 2  &  5 & 6  \\
\hline
\end{tabular}

\begin{tabular}{c}
$\vert S_4^0 \rangle$=
\end{tabular}
\begin{tabular}{|c|c|c|}
\hline
                1   & 2   & 5    \\
\cline{1-3} 3  &  4 & 6  \\
\hline
\end{tabular}
\begin{tabular}{c}
$\vert S_5^0 \rangle$=
\end{tabular}
\begin{tabular}{|c|c|c|}
\hline
                1   & 3   & 5    \\
\cline{1-3} 2  &  4 & 6  \\
\hline
\end{tabular}
\item S=1 : 9 basis functions with Young tableau [4,2]
\begin{tabular}{c}
$\vert S_1^1 \rangle$=
\end{tabular}
\begin{tabular}{|c|c|c|c|}
\hline
                  1 & 2 & 3 & 4   \\
\cline{1-4}
\multicolumn{1}{|c|}{5} & \multicolumn{1}{c|}{6}  \\
\cline{1-2}
\end{tabular}
\begin{tabular}{c}
$\vert S_2^1 \rangle$=
\end{tabular}
\begin{tabular}{|c|c|c|c|}
\hline
                  1 & 2 & 3 & 5   \\
\cline{1-4}
\multicolumn{1}{|c|}{4} & \multicolumn{1}{c|}{6}  \\
\cline{1-2}
\end{tabular}
\begin{tabular}{c}
$\vert S_3^1 \rangle$=
\end{tabular}
\begin{tabular}{|c|c|c|c|}
\hline
                  1 & 2 & 4 & 5   \\
\cline{1-4}
\multicolumn{1}{|c|}{3} & \multicolumn{1}{c|}{6}  \\
\cline{1-2}
\end{tabular}

\begin{tabular}{c}
$\vert S_4^1 \rangle$=
\end{tabular}
\begin{tabular}{|c|c|c|c|}
\hline
                  1 & 3 & 4 & 5   \\
\cline{1-4}
\multicolumn{1}{|c|}{2} & \multicolumn{1}{c|}{6}  \\
\cline{1-2}
\end{tabular}
\begin{tabular}{c}
$\vert S_5^1 \rangle$=
\end{tabular}
\begin{tabular}{|c|c|c|c|}
\hline
                  1 & 2 & 3 & 6   \\
\cline{1-4}
\multicolumn{1}{|c|}{4} & \multicolumn{1}{c|}{5}  \\
\cline{1-2}
\end{tabular}
\begin{tabular}{c}
$\vert S_6^1 \rangle$=
\end{tabular}
\begin{tabular}{|c|c|c|c|}
\hline
                  1 & 2 & 4 & 6   \\
\cline{1-4}
\multicolumn{1}{|c|}{3} & \multicolumn{1}{c|}{5}  \\
\cline{1-2}
\end{tabular}

\begin{tabular}{c}
$\vert S_7^1 \rangle$=
\end{tabular}
\begin{tabular}{|c|c|c|c|}
\hline
                  1 & 3 & 4 & 6   \\
\cline{1-4}
\multicolumn{1}{|c|}{2} & \multicolumn{1}{c|}{5}  \\
\cline{1-2}
\end{tabular}
\begin{tabular}{c}
$\vert S_8^1 \rangle$=
\end{tabular}
\begin{tabular}{|c|c|c|c|}
\hline
                  1 & 2 & 5 & 6   \\
\cline{1-4}
\multicolumn{1}{|c|}{3} & \multicolumn{1}{c|}{4}  \\
\cline{1-2}
\end{tabular}
\begin{tabular}{c}
$\vert S_9^1 \rangle$=
\end{tabular}
\begin{tabular}{|c|c|c|c|}
\hline
                  1 & 3 & 5 & 6   \\
\cline{1-4}
\multicolumn{1}{|c|}{2} & \multicolumn{1}{c|}{4}  \\
\cline{1-2}
\end{tabular}
\item S=2 : 5 basis functions with Young tableau [5,1] 
\begin{tabular}{c}
$\vert S_1^2 \rangle$=
\end{tabular}
\begin{tabular}{|c|c|c|c|c|}
\hline
                  1 & 2 & 3 & 4 & 5   \\
\cline{1-5}
\multicolumn{1}{|c|}{6}  \\
\cline{1-1}
\end{tabular}
\begin{tabular}{c}
$\vert S_2^2 \rangle$=
\end{tabular}
\begin{tabular}{|c|c|c|c|c|}
\hline
                  1 & 2 & 3 & 4 & 6   \\
\cline{1-5}
\multicolumn{1}{|c|}{5}  \\
\cline{1-1}
\end{tabular}

\begin{tabular}{c}
$\vert S_3^2 \rangle$=
\end{tabular}
\begin{tabular}{|c|c|c|c|c|}
\hline
                  1 & 2 & 3 & 5 & 6   \\
\cline{1-5}
\multicolumn{1}{|c|}{4}  \\
\cline{1-1}
\end{tabular}
\begin{tabular}{c}
$\vert S_4^2 \rangle$=
\end{tabular}
\begin{tabular}{|c|c|c|c|c|}
\hline
                  1 & 2 & 4 & 5 & 6   \\
\cline{1-5}
\multicolumn{1}{|c|}{3}  \\
\cline{1-1}
\end{tabular}

\begin{tabular}{c}
$\vert S_5^2 \rangle$=
\end{tabular}
\begin{tabular}{|c|c|c|c|c|}
\hline
                  1 & 3 & 4 & 5 & 6   \\
\cline{1-5}
\multicolumn{1}{|c|}{2}  \\
\cline{1-1}
\end{tabular}
\item S=3 : 1 basis function with Young tableau [6] 
\begin{tabular}{c}
$\vert S^3 \rangle$=
\end{tabular}
\begin{tabular}{|c|c|c|c|c|c|}
\hline
                  1 & 2 & 3 & 4 & 5 & 6   \\
\hline
\end{tabular}
\end{itemize}

We are now in a position to construct the completely antisymmetric function of the color $\otimes$ isospin $\otimes$ spin state of dibaryon, which we will denote by $\vert C_i, I_j, S_k \rangle$. 
In particular, we choose the $\vert I, S \rangle$ basis function to be in the [3,3] representation, and in the  conjugate of the color singlet [2,2,2], so that the color singlet $\otimes$  isospin $\otimes$ spins becomes fully antisymmetric.
 With this IS coupling scheme, we can find the fully antisymmetric function of $\vert C, I, S \rangle$ for all of (I,S), by using the Clebsch-Gordan (CG) coefficient for making the representation of [3,3] of isospin $\otimes$ spin function. In calculating the Clebsch-Gordan coefficients of the element of the permutation group, $S_6$, it is convenient to use the factorization property that factorizes the CG coefficients of $S_n$ into an isoscalar factor,   which is called K matrix, and a CG coefficients of $S_{n-1}$.  For example, the isoscalar factor can be defined by~\cite{Stancu:1991rc},
\begin{align}
&S([f^{\prime}]p^{\prime}q^{\prime}y^{\prime}[f^{\prime\prime}]p^{\prime\prime}q^{\prime\prime}y^{\prime\prime}\vert[f]pqy)=\nonumber\\
&K([f^{\prime}]p^{\prime}[f^{\prime\prime}]p^{\prime\prime}\vert[f]p)
S([f^{\prime}_{p^{\prime}}]q^{\prime}y^{\prime}[f^{\prime\prime}_{p^{\prime\prime}}]q^{\prime\prime}y^{\prime\prime}\vert[f_p]qy),
\end{align}
where S in the left-hand (right-hand) side is a CG coefficients of $S_n$ ($S_{n-1}$).  In this notation,
$[f_p]$ is the Young tableau associated to $S_{n-1}$ which can be obtained from $[f]$, the Young tableau of $S_n$, by removing the n-th particle characterized by $pqy$; where 
$p$ is the position of the n-th particle in the row, $q$, the position of the (n-1)-th particle in the row, $y$, the position of the (n-2)-th particle in the row, respectively. In our case, by repeating the process of factorizing the CG coefficients of $S_6$  further, we find the CG coefficients from the following formula,
\begin{align}                                     
&S([f^{\prime}]p^{\prime}q^{\prime}y^{\prime}r^{\prime}[f^{\prime\prime}]p^{\prime\prime}q^{\prime\prime}y^{\prime\prime}r^{\prime\prime}\vert[f]pqyr)=\nonumber\\
&K([f^{\prime}]p^{\prime}[f^{\prime\prime}]p^{\prime\prime}\vert[f]p) K([f^{\prime}_{p^{\prime}}]q^{\prime}[f^{\prime\prime}_{p^{\prime\prime}}]q^{\prime\prime}\vert[f_p]q) \nonumber\\
&K([f^{\prime}_{p^{\prime}q^{\prime}}]y^{\prime}[f^{\prime\prime}_{p^{\prime\prime}q^{\prime\prime}}]y^{\prime\prime}\vert[f_{pq}]y)
S([f^{\prime}_{p^{\prime}q^{\prime}y^{\prime}}]r^{\prime}[f^{\prime\prime}_{p^{\prime\prime}q^{\prime\prime}y^{\prime\prime}}]r^{\prime\prime}\vert[f_{pqy}]r),
\end{align}
where S in the third row is the CG coefficient of $S_3$. When we calculate the CG coefficients, we use the relevant isoscalar factors for $S_4$, $S_5$, and $S_6$ in Eq.~(23) which is obtained by Stancu and Pepin~\cite{Stancu:1999qr}.   In the case of (I,S)=(0,1), the $\vert [I^0, S^1] \rangle$ basis functions belonging to the Young tableau of [3,3] are presented as the followings :
\begin{align}
&\begin{tabular}{c}
$\vert [I^0, S^1]_1 \rangle$=
\end{tabular}
\begin{tabular}{|c|c|c|}
\hline
                1   & 2   & 3    \\
\cline{1-3} 4  &  5 & 6  \\
\hline
\end{tabular}
=\nonumber
\\
&\frac{\sqrt{6}}{9}\vert I^0_1 \rangle \otimes \vert S^1_1 \rangle+\frac{\sqrt{10}}{9}\vert I^0_1 \rangle \otimes \vert S^1_2 \rangle+\frac{2\sqrt{5}}{9}\vert I^0_1 \rangle \otimes \vert S^1_5 \rangle- \nonumber\\ 
&\frac{\sqrt{10}}{18}\vert I^0_2 \rangle \otimes \vert S^1_3 \rangle-\frac{\sqrt{5}}{9}\vert I^0_2 \rangle \otimes \vert S^1_6 \rangle+\frac{\sqrt{60}}{36}\vert I^0_2 \rangle \otimes \vert S^1_8 \rangle- \nonumber\\ 
&\frac{\sqrt{10}}{18}\vert I^0_3 \rangle \otimes \vert S^1_4 \rangle-\frac{\sqrt{5}}{9}\vert I^0_3 \rangle \otimes \vert S^1_7 \rangle+\frac{\sqrt{60}}{36}\vert I^0_3 \rangle \otimes \vert S^1_9 \rangle- \nonumber\\
&\frac{\sqrt{30}}{18}\vert I^0_4 \rangle \otimes \vert S^1_3 \rangle+\frac{\sqrt{60}}{36}\vert I^0_4 \rangle \otimes \vert S^1_6 \rangle-\frac{\sqrt{30}}{18}\vert I^0_5 \rangle \otimes \vert S^1_4 \rangle+\nonumber\\ 
&\frac{\sqrt{60}}{36}\vert I^0_5 \rangle \otimes \vert S^1_7 \rangle.
\end{align}

\begin{align}
&\begin{tabular}{c}
$\vert [I^0, S^1]_2 \rangle$=
\end{tabular}
\begin{tabular}{|c|c|c|}
\hline
                1   & 2   & 4    \\
\cline{1-3} 3  &  5 & 6  \\
\hline
\end{tabular}
=\nonumber
\\
&-\frac{\sqrt{10}}{18}\vert I^0_1 \rangle \otimes \vert S^1_3 \rangle-\frac{\sqrt{20}}{18}\vert I^0_1 \rangle \otimes \vert S^1_6 \rangle+\frac{\sqrt{15}}{18}\vert I^0_1 \rangle \otimes \vert S^1_8 \rangle \nonumber\\ 
&+\frac{\sqrt{6}}{9}\vert I^0_2 \rangle \otimes \vert S^1_1 \rangle-\frac{\sqrt{10}}{18}\vert I^0_2 \rangle \otimes \vert S^1_2 \rangle+\frac{\sqrt{5}}{9}\vert I^0_2 \rangle \otimes \vert S^1_4 \rangle \nonumber\\ 
&-\frac{\sqrt{20}}{18}\vert I^0_2 \rangle \otimes \vert S^1_5 \rangle+\frac{\sqrt{10}}{9}\vert I^0_2 \rangle \otimes \vert S^1_6 \rangle+\frac{\sqrt{30}}{36}\vert I^0_2 \rangle \otimes \vert S^1_8 \rangle  \nonumber\\
&+\frac{\sqrt{5}}{9}\vert I^0_3 \rangle \otimes \vert S^1_4 \rangle+\frac{\sqrt{10}}{9}\vert I^0_3 \rangle \otimes \vert S^1_7 \rangle+\frac{\sqrt{30}}{36}\vert I^0_3 \rangle \otimes \vert S^1_9 \rangle \nonumber\\ 
&-\frac{\sqrt{30}}{18}\vert I^0_4 \rangle \otimes \vert S^1_2 \rangle-\frac{\sqrt{15}}{18}\vert I^0_4 \rangle \otimes \vert S^1_3 \rangle+\frac{\sqrt{15}}{18}\vert I^0_4 \rangle \otimes \vert S^1_5 \rangle \nonumber\\  
&+\frac{\sqrt{30}}{36}\vert I^0_4 \rangle \otimes \vert S^1_6 \rangle+\frac{\sqrt{15}}{18}\vert I^0_5 \rangle \otimes \vert S^1_4 \rangle-\frac{\sqrt{30}}{36}\vert I^0_5 \rangle \otimes \vert S^1_7 \rangle.
\end{align}
\begin{align}
&\begin{tabular}{c}
$\vert [I^0, S^1]_3 \rangle$=
\end{tabular}
\begin{tabular}{|c|c|c|}
\hline
                1   & 3   & 4    \\
\cline{1-3} 2  &  5 & 6  \\
\hline
\end{tabular}
=\nonumber
\\
&-\frac{\sqrt{10}}{18}\vert I^0_1 \rangle \otimes \vert S^1_4 \rangle-\frac{\sqrt{20}}{18}\vert I^0_1 \rangle \otimes \vert S^1_7 \rangle+\frac{\sqrt{15}}{18}\vert I^0_1 \rangle \otimes \vert S^1_9 \rangle \nonumber\\ 
&-\frac{\sqrt{5}}{9}\vert I^0_2 \rangle \otimes \vert S^1_4 \rangle-\frac{\sqrt{10}}{9}\vert I^0_2 \rangle \otimes \vert S^1_7 \rangle-\frac{\sqrt{30}}{36}\vert I^0_2 \rangle \otimes \vert S^1_9 \rangle \nonumber\\ 
&+\frac{\sqrt{6}}{9}\vert I^0_3 \rangle \otimes \vert S^1_1 \rangle-\frac{\sqrt{10}}{18}\vert I^0_3 \rangle \otimes \vert S^1_2 \rangle-\frac{\sqrt{20}}{18}\vert I^0_3 \rangle \otimes \vert S^1_5 \rangle  \nonumber\\
&-\frac{\sqrt{5}}{9}\vert I^0_3 \rangle \otimes \vert S^1_3 \rangle-\frac{\sqrt{10}}{9}\vert I^0_3 \rangle \otimes \vert S^1_6 \rangle-\frac{\sqrt{30}}{36}\vert I^0_3 \rangle \otimes \vert S^1_8 \rangle \nonumber\\ 
&+\frac{\sqrt{15}}{18}\vert I^0_4 \rangle \otimes \vert S^1_4 \rangle-\frac{\sqrt{30}}{36}\vert I^0_4 \rangle \otimes \vert S^1_7 \rangle-\frac{\sqrt{30}}{18}\vert I^0_5 \rangle \otimes \vert S^1_2 \rangle \nonumber\\  
&+\frac{\sqrt{15}}{18}\vert I^0_5 \rangle \otimes \vert S^1_4 \rangle+\frac{\sqrt{15}}{18}\vert I^0_5 \rangle \otimes \vert S^1_5 \rangle+\frac{\sqrt{30}}{36}\vert I^0_5 \rangle \otimes \vert S^1_6 \rangle.
\end{align}
\begin{align}
&\begin{tabular}{c}
$\vert [I^0, S^1]_4 \rangle$=
\end{tabular}
\begin{tabular}{|c|c|c|}
\hline
                1   & 2   & 5    \\
\cline{1-3} 3  &  4 & 6  \\
\hline
\end{tabular}
=\nonumber
\\
&-\frac{\sqrt{30}}{18}\vert I^0_1 \rangle \otimes \vert S^1_3 \rangle+\frac{\sqrt{15}}{18}\vert I^0_1 \rangle \otimes \vert S^1_6 \rangle-\frac{\sqrt{30}}{18}\vert I^0_2 \rangle \otimes \vert S^1_2 \rangle \nonumber\\ 
&-\frac{\sqrt{15}}{18}\vert I^0_2 \rangle \otimes \vert S^1_3 \rangle+\frac{\sqrt{15}}{18}\vert I^0_2 \rangle \otimes \vert S^1_5 \rangle+\frac{\sqrt{30}}{36}\vert I^0_2 \rangle \otimes \vert S^1_6 \rangle \nonumber\\ 
&-\frac{\sqrt{15}}{18}\vert I^0_3 \rangle \otimes \vert S^1_4 \rangle+\frac{\sqrt{30}}{36}\vert I^0_3 \rangle \otimes \vert S^1_7 \rangle-\frac{\sqrt{6}}{6}\vert I^0_4 \rangle \otimes \vert S^1_1 \rangle  \nonumber\\
&+\frac{\sqrt{30}}{12}\vert I^0_4 \rangle \otimes \vert S^1_8 \rangle-\frac{\sqrt{30}}{12}\vert I^0_5 \rangle \otimes \vert S^1_9 \rangle.
\end{align}
\begin{align}
&\begin{tabular}{c}
$\vert [I^0, S^1]_5 \rangle$=
\end{tabular}
\begin{tabular}{|c|c|c|}
\hline
                1   & 3   & 5    \\
\cline{1-3} 2  &  4 & 6  \\
\hline
\end{tabular}
=\nonumber
\\
&-\frac{\sqrt{30}}{18}\vert I^0_1 \rangle \otimes \vert S^1_4 \rangle+\frac{\sqrt{15}}{18}\vert I^0_1 \rangle \otimes \vert S^1_7 \rangle-\frac{\sqrt{15}}{18}\vert I^0_2 \rangle \otimes \vert S^1_4 \rangle \nonumber\\ 
&+\frac{\sqrt{30}}{36}\vert I^0_2 \rangle \otimes \vert S^1_7 \rangle-\frac{\sqrt{30}}{18}\vert I^0_3 \rangle \otimes \vert S^1_2 \rangle+\frac{\sqrt{15}}{18}\vert I^0_3 \rangle \otimes \vert S^1_3 \rangle \nonumber\\ 
&+\frac{\sqrt{15}}{18}\vert I^0_3 \rangle \otimes \vert S^1_5 \rangle-\frac{\sqrt{30}}{36}\vert I^0_3 \rangle \otimes \vert S^1_6 \rangle-\frac{\sqrt{30}}{12}\vert I^0_4 \rangle \otimes \vert S^1_9 \rangle  \nonumber\\
&-\frac{\sqrt{6}}{6}\vert I^0_5 \rangle \otimes \vert S^1_1 \rangle-\frac{\sqrt{30}}{12}\vert I^0_5 \rangle \otimes \vert S^1_8 \rangle.
\end{align}
Coupling the isospin $\otimes$ spin basis function obtained from the IS scheme to the color singlet basis function, we find 
the color $\otimes$ isospin $\otimes$ spin state satisfying the fully antisymmetry property for (I,S)=(0,1).  This  is given by 
\begin{align}
\vert C, I^0, S^1 \rangle=\frac{1}{\sqrt{5}}(&\vert C_1 \rangle \otimes \vert [I^0, S^1]_5 \rangle-\vert C_2\rangle                           \otimes \vert [I^0, S^1]_4 \rangle-  \nonumber\\                                                                                                                     
&\vert C_3 \rangle \otimes \vert [I^0, S^1]_3 \rangle+\vert C_4 \rangle \otimes \vert [I^0, S^1]_2 \rangle-    \nonumber\\                                                                                                                     
&\vert C_5 \rangle \otimes \vert [I^0, S^1]_1 \rangle).
\end{align}

In the case of (I,S)=(1,0), the $\vert [I^1, S^0] \rangle$ basis
functions belonging to the Young tableau of [3,3] are presented as the followings : 

\begin{align}
&\begin{tabular}{c}
$\vert [I^1, S^0]_1 \rangle$=
\end{tabular}
\begin{tabular}{|c|c|c|}
\hline
                1   & 2   & 3    \\
\cline{1-3} 4  &  5 & 6  \\
\hline
\end{tabular}
=\nonumber
\\
&\frac{\sqrt{6}}{9}\vert I^1_1 \rangle \otimes \vert S^0_1 \rangle+\frac{\sqrt{10}}{9}\vert I^1_2 \rangle \otimes \vert S^0_1 \rangle+\frac{2\sqrt{5}}{9}\vert I^1_5 \rangle \otimes \vert S^0_1 \rangle- \nonumber\\ 
&\frac{\sqrt{10}}{18}\vert I^1_3 \rangle \otimes \vert S^0_2 \rangle-\frac{\sqrt{5}}{9}\vert I^1_6 \rangle \otimes \vert S^0_2 \rangle+\frac{\sqrt{60}}{36}\vert I^1_8 \rangle \otimes \vert S^0_2 \rangle- \nonumber\\ 
&\frac{\sqrt{10}}{18}\vert I^1_4 \rangle \otimes \vert S^0_3 \rangle-\frac{\sqrt{5}}{9}\vert I^1_7 \rangle \otimes \vert S^0_3 \rangle+\frac{\sqrt{60}}{36}\vert I^1_9 \rangle \otimes \vert S^0_3 \rangle- \nonumber\\
&\frac{\sqrt{30}}{18}\vert I^1_3 \rangle \otimes \vert S^0_4 \rangle+\frac{\sqrt{60}}{36}\vert I^1_6 \rangle \otimes \vert S^0_4 \rangle-\frac{\sqrt{30}}{18}\vert I^1_4 \rangle \otimes \vert S^0_5 \rangle+\nonumber\\ 
&\frac{\sqrt{60}}{36}\vert I^1_7 \rangle \otimes \vert S^0_5 \rangle.
\end{align}
\begin{align}
&\begin{tabular}{c}
$\vert [I^1, S^0]_2 \rangle$=
\end{tabular}
\begin{tabular}{|c|c|c|}
\hline
                1   & 2   & 4    \\
\cline{1-3} 3  &  5 & 6  \\
\hline
\end{tabular}
=\nonumber
\\
&-\frac{\sqrt{10}}{18}\vert I^1_3 \rangle \otimes \vert S^0_1 \rangle-\frac{\sqrt{20}}{18}\vert I^1_6 \rangle \otimes \vert S^0_1 \rangle+\frac{\sqrt{15}}{18}\vert I^1_8 \rangle \otimes \vert S^0_1 \rangle \nonumber\\ 
&+\frac{\sqrt{6}}{9}\vert I^1_1 \rangle \otimes \vert S^0_2 \rangle-\frac{\sqrt{10}}{18}\vert I^1_2 \rangle \otimes \vert S^0_2 \rangle+\frac{\sqrt{5}}{9}\vert I^1_4 \rangle \otimes \vert S^0_2 \rangle \nonumber\\ 
&-\frac{\sqrt{20}}{18}\vert I^1_5 \rangle \otimes \vert S^0_2 \rangle+\frac{\sqrt{10}}{9}\vert I^1_6 \rangle \otimes \vert S^0_2 \rangle+\frac{\sqrt{30}}{36}\vert I^1_8 \rangle \otimes \vert S^0_2 \rangle  \nonumber\\
&+\frac{\sqrt{5}}{9}\vert I^1_4 \rangle \otimes \vert S^0_3 \rangle+\frac{\sqrt{10}}{9}\vert I^1_7 \rangle \otimes \vert S^0_3 \rangle+\frac{\sqrt{30}}{36}\vert I^1_9 \rangle \otimes \vert S^0_3 \rangle \nonumber\\ 
&-\frac{\sqrt{30}}{18}\vert I^1_2 \rangle \otimes \vert S^0_4 \rangle-\frac{\sqrt{15}}{18}\vert I^1_3 \rangle \otimes \vert S^0_4 \rangle+\frac{\sqrt{15}}{18}\vert I^1_5 \rangle \otimes \vert S^0_4 \rangle \nonumber\\  
&+\frac{\sqrt{30}}{36}\vert I^1_6 \rangle \otimes \vert S^0_4 \rangle+\frac{\sqrt{15}}{18}\vert I^1_4 \rangle \otimes \vert S^0_5 \rangle-\frac{\sqrt{30}}{36}\vert I^1_7 \rangle \otimes \vert S^0_5 \rangle.
\end{align}
\begin{align}
&\begin{tabular}{c}
$\vert [I^1, S^0]_3 \rangle$=
\end{tabular}
\begin{tabular}{|c|c|c|}
\hline
                1   & 3   & 4    \\
\cline{1-3} 2  &  5 & 6  \\
\hline
\end{tabular}
=\nonumber
\\
&-\frac{\sqrt{10}}{18}\vert I^1_4 \rangle \otimes \vert S^0_1 \rangle-\frac{\sqrt{20}}{18}\vert I^1_7 \rangle \otimes \vert S^0_1 \rangle+\frac{\sqrt{15}}{18}\vert I^1_9 \rangle \otimes \vert S^0_1 \rangle \nonumber\\ 
&-\frac{\sqrt{5}}{9}\vert I^1_4 \rangle \otimes \vert S^0_2 \rangle-\frac{\sqrt{10}}{9}\vert I^1_7 \rangle \otimes \vert S^0_2 \rangle-\frac{\sqrt{30}}{36}\vert I^1_9 \rangle \otimes \vert S^0_2 \rangle \nonumber\\ 
&+\frac{\sqrt{6}}{9}\vert I^1_1 \rangle \otimes \vert S^0_3 \rangle-\frac{\sqrt{10}}{18}\vert I^1_2 \rangle \otimes \vert S^0_3 \rangle-\frac{\sqrt{20}}{18}\vert I^1_5 \rangle \otimes \vert S^0_3 \rangle  \nonumber\\
&-\frac{\sqrt{5}}{9}\vert I^1_3 \rangle \otimes \vert S^0_3 \rangle-\frac{\sqrt{10}}{9}\vert I^1_6 \rangle \otimes \vert S^0_3 \rangle-\frac{\sqrt{30}}{36}\vert I^1_8 \rangle \otimes \vert S^0_3 \rangle \nonumber\\ 
&+\frac{\sqrt{15}}{18}\vert I^1_4 \rangle \otimes \vert S^0_4 \rangle-\frac{\sqrt{30}}{36}\vert I^1_7 \rangle \otimes \vert S^0_4 \rangle-\frac{\sqrt{30}}{18}\vert I^1_2 \rangle \otimes \vert S^0_5 \rangle \nonumber\\  
&+\frac{\sqrt{15}}{18}\vert I^1_4 \rangle \otimes \vert S^0_5 \rangle+\frac{\sqrt{15}}{18}\vert I^1_5 \rangle \otimes \vert S^0_5 \rangle+\frac{\sqrt{30}}{36}\vert I^1_6 \rangle \otimes \vert S^0_5 \rangle.
\end{align}
\begin{align}
&\begin{tabular}{c}
$\vert [I^1, S^0]_4 \rangle$=
\end{tabular}
\begin{tabular}{|c|c|c|}
\hline
                1   & 2   & 5    \\
\cline{1-3} 3  &  4 & 6  \\
\hline
\end{tabular}
=\nonumber
\\
&-\frac{\sqrt{30}}{18}\vert I^1_3 \rangle \otimes \vert S^0_1 \rangle+\frac{\sqrt{15}}{18}\vert I^1_6 \rangle \otimes \vert S^0_1 \rangle-\frac{\sqrt{30}}{18}\vert I^1_2 \rangle \otimes \vert S^0_2 \rangle \nonumber\\ 
&-\frac{\sqrt{15}}{18}\vert I^1_3 \rangle \otimes \vert S^0_2 \rangle+\frac{\sqrt{15}}{18}\vert I^1_5 \rangle \otimes \vert S^0_2 \rangle+\frac{\sqrt{30}}{36}\vert I^1_6 \rangle \otimes \vert S^0_2 \rangle \nonumber\\ 
&-\frac{\sqrt{15}}{18}\vert I^1_4 \rangle \otimes \vert S^0_3 \rangle+\frac{\sqrt{30}}{36}\vert I^1_7 \rangle \otimes \vert S^0_3 \rangle-\frac{\sqrt{6}}{6}\vert I^1_1 \rangle \otimes \vert S^0_4 \rangle  \nonumber\\
&+\frac{\sqrt{30}}{12}\vert I^1_8 \rangle \otimes \vert S^0_4 \rangle-\frac{\sqrt{30}}{12}\vert I^1_9 \rangle \otimes \vert S^0_5 \rangle.
\end{align}
\begin{align}
&\begin{tabular}{c}
$\vert [I^1, S^0]_5 \rangle$=
\end{tabular}
\begin{tabular}{|c|c|c|}
\hline
                1   & 3   & 5    \\
\cline{1-3} 2  &  4 & 6  \\
\hline
\end{tabular}
=\nonumber
\\
&-\frac{\sqrt{30}}{18}\vert I^1_4 \rangle \otimes \vert S^0_1 \rangle+\frac{\sqrt{15}}{18}\vert I^1_7 \rangle \otimes \vert S^0_1 \rangle-\frac{\sqrt{15}}{18}\vert I^1_4 \rangle \otimes \vert S^0_2 \rangle \nonumber\\ 
&+\frac{\sqrt{30}}{36}\vert I^1_7 \rangle \otimes \vert S^0_2 \rangle-\frac{\sqrt{30}}{18}\vert I^1_2 \rangle \otimes \vert S^0_3 \rangle+\frac{\sqrt{15}}{18}\vert I^1_3 \rangle \otimes \vert S^0_3 \rangle \nonumber\\ 
&+\frac{\sqrt{15}}{18}\vert I^1_5 \rangle \otimes \vert S^0_3 \rangle-\frac{\sqrt{30}}{36}\vert I^1_6 \rangle \otimes \vert S^0_3 \rangle-\frac{\sqrt{30}}{12}\vert I^1_9 \rangle \otimes \vert S^0_4 \rangle  \nonumber\\
&-\frac{\sqrt{6}}{6}\vert I^1_1 \rangle \otimes \vert S^0_5 \rangle-\frac{\sqrt{30}}{12}\vert I^1_8 \rangle \otimes \vert S^0_5 \rangle.
\end{align}
Likewise, we find the color $\otimes$ isospin $\otimes$ spin state satisfying the fully antisymmetry property for (I,S)=(1,0) to be given by 
\begin{align}
\vert C, I^1, S^0 \rangle=\frac{1}{\sqrt{5}}(&\vert C_1 \rangle \otimes \vert [I^1, S^0]_5 \rangle-\vert C_2\rangle                           \otimes \vert [I^1, S^0]_4 \rangle-  \nonumber\\                                                                                                                     
&\vert C_3 \rangle \otimes \vert [I^1, S^0]_3 \rangle+\vert C_4 \rangle \otimes \vert [I^1, S^0]_2 \rangle-    \nonumber\\                                                                                                                     
&\vert C_5 \rangle \otimes \vert [I^1, S^0]_1 \rangle).
\end{align}

In the case of (I,S)=(1,2), the $\vert [I^1, S^2] \rangle$ basis
functions belonging to the Young tableau of [3,3] are presented as the followings :

\begin{align}
&\begin{tabular}{c}
$\vert [I^1, S^2]_1 \rangle$=
\end{tabular}
\begin{tabular}{|c|c|c|}
\hline
                1   & 2   & 3    \\
\cline{1-3} 4  &  5 & 6  \\
\hline
\end{tabular}
=\nonumber
\\
&\frac{\sqrt{15}}{9}\vert I^1_1 \rangle \otimes \vert S^2_3 \rangle+\frac{\sqrt{15}}{9}\vert I^1_2 \rangle \otimes \vert S^2_2 \rangle+\frac{2}{9}\vert I^1_2 \rangle \otimes \vert S^2_3 \rangle- \nonumber\\ 
&\frac{1}{9}\vert I^1_3 \rangle \otimes \vert S^2_4 \rangle-\frac{1}{9}\vert I^1_4 \rangle \otimes \vert S^2_5 \rangle+\frac{\sqrt{5}}{5}\vert I^1_5 \rangle \otimes \vert S^2_1 \rangle+ \nonumber\\ 
&\frac{\sqrt{120}}{45}\vert I^1_5 \rangle \otimes \vert S^2_2 \rangle+\frac{2\sqrt{2}}{9}\vert I^1_5 \rangle \otimes \vert S^2_3 \rangle-\frac{\sqrt{2}}{9}\vert I^1_6 \rangle \otimes \vert S^2_4 \rangle- \nonumber\\
&\frac{\sqrt{2}}{9}\vert I^1_7 \rangle \otimes \vert S^2_5 \rangle-\frac{\sqrt{6}}{9}\vert I^1_8 \rangle \otimes \vert S^2_4 \rangle-\frac{\sqrt{6}}{9}\vert I^1_9 \rangle \otimes \vert S^2_5 \rangle.
\end{align}
\begin{align}
&\begin{tabular}{c}
$\vert [I^1, S^2]_2 \rangle$=
\end{tabular}
\begin{tabular}{|c|c|c|}
\hline
                1   & 2   & 4    \\
\cline{1-3} 3  &  5 & 6  \\
\hline
\end{tabular}
=\nonumber
\\
&\frac{\sqrt{15}}{9}\vert I^1_1 \rangle \otimes \vert S^2_4 \rangle-\frac{1}{9}\vert I^1_2 \rangle \otimes \vert S^2_4 \rangle+\frac{\sqrt{15}}{9}\vert I^1_3 \rangle \otimes \vert S^2_2 \rangle- \nonumber\\ 
&\frac{1}{9}\vert I^1_3 \rangle \otimes \vert S^2_3 \rangle+\frac{\sqrt{2}}{9}\vert I^1_3 \rangle \otimes \vert S^2_4 \rangle-\frac{\sqrt{2}}{9}\vert I^1_4 \rangle \otimes \vert S^2_5 \rangle- \nonumber\\ 
&\frac{\sqrt{2}}{9}\vert I^1_5 \rangle \otimes \vert S^2_4 \rangle+\frac{\sqrt{5}}{5}\vert I^1_6 \rangle \otimes \vert S^2_1 \rangle+\frac{\sqrt{120}}{45}\vert I^1_6 \rangle \otimes \vert S^2_2 \rangle- \nonumber\\
&\frac{\sqrt{2}}{9}\vert I^1_6 \rangle \otimes \vert S^2_3 \rangle+\frac{2}{9}\vert I^1_6 \rangle \otimes \vert S^2_4 \rangle-\frac{2}{9}\vert I^1_7 \rangle \otimes \vert S^2_5 \rangle- \nonumber\\
&\frac{\sqrt{6}}{9}\vert I^1_8 \rangle \otimes \vert S^2_3 \rangle-\frac{\sqrt{3}}{9}\vert I^1_8 \rangle \otimes \vert S^2_4 \rangle+\frac{\sqrt{3}}{9}\vert I^1_9 \rangle \otimes \vert S^2_5 \rangle.
\end{align}
\begin{align}
&\begin{tabular}{c}
$\vert [I^1, S^2]_3 \rangle$=
\end{tabular}
\begin{tabular}{|c|c|c|}
\hline
                1   & 3   & 4    \\
\cline{1-3} 2  &  5 & 6  \\
\hline
\end{tabular}
=\nonumber
\\
&\frac{\sqrt{15}}{9}\vert I^1_1 \rangle \otimes \vert S^2_5 \rangle-\frac{1}{9}\vert I^1_2 \rangle \otimes \vert S^2_5 \rangle-\frac{\sqrt{2}}{9}\vert I^1_3 \rangle \otimes \vert S^2_5 \rangle+ \nonumber\\ 
&\frac{\sqrt{15}}{9}\vert I^1_4 \rangle \otimes \vert S^2_2 \rangle-\frac{1}{9}\vert I^1_4 \rangle \otimes \vert S^2_3 \rangle-\frac{\sqrt{2}}{9}\vert I^1_4 \rangle \otimes \vert S^2_4 \rangle- \nonumber\\ 
&\frac{\sqrt{2}}{9}\vert I^1_5 \rangle \otimes \vert S^2_5 \rangle-\frac{2}{9}\vert I^1_6 \rangle \otimes \vert S^2_5 \rangle+\frac{\sqrt{5}}{5}\vert I^1_7 \rangle \otimes \vert S^2_1 \rangle+ \nonumber\\
&\frac{\sqrt{120}}{45}\vert I^1_7 \rangle \otimes \vert S^2_2 \rangle-\frac{\sqrt{2}}{9}\vert I^1_7 \rangle \otimes \vert S^2_3 \rangle-\frac{2}{9}\vert I^1_7 \rangle \otimes \vert S^2_4 \rangle+ \nonumber\\
&\frac{\sqrt{3}}{9}\vert I^1_8 \rangle \otimes \vert S^2_5 \rangle-\frac{\sqrt{6}}{9}\vert I^1_9 \rangle \otimes \vert S^2_3 \rangle+\frac{\sqrt{3}}{9}\vert I^1_9 \rangle \otimes \vert S^2_4 \rangle.
\end{align}
\begin{align}
&\begin{tabular}{c}
$\vert [I^1, S^2]_4 \rangle$=
\end{tabular}
\begin{tabular}{|c|c|c|}
\hline
                1   & 2   & 5    \\
\cline{1-3} 3  &  4 & 6  \\
\hline
\end{tabular}
=\nonumber
\\
&\frac{2\sqrt{3}}{9}\vert I^1_2 \rangle \otimes \vert S^2_4 \rangle+\frac{2\sqrt{3}}{9}\vert I^1_3 \rangle \otimes \vert S^2_3 \rangle+\frac{\sqrt{6}}{9}\vert I^1_3 \rangle \otimes \vert S^2_4 \rangle- \nonumber\\ 
&\frac{\sqrt{6}}{9}\vert I^1_4 \rangle \otimes \vert S^2_5 \rangle-\frac{\sqrt{6}}{9}\vert I^1_5 \rangle \otimes \vert S^2_4 \rangle-\frac{\sqrt{6}}{9}\vert I^1_6 \rangle \otimes \vert S^2_3 \rangle- \nonumber\\ 
&\frac{\sqrt{3}}{9}\vert I^1_6 \rangle \otimes \vert S^2_4 \rangle+\frac{\sqrt{3}}{9}\vert I^1_7 \rangle \otimes \vert S^2_5 \rangle+\frac{\sqrt{5}}{5}\vert I^1_8 \rangle \otimes \vert S^2_1 \rangle- \nonumber\\
&\frac{\sqrt{270}}{45}\vert I^1_8 \rangle \otimes \vert S^2_2 \rangle.
\end{align}
\begin{align}
&\begin{tabular}{c}
$\vert [I^1, S^2]_5 \rangle$=
\end{tabular}
\begin{tabular}{|c|c|c|}
\hline
                1   & 3   & 5    \\
\cline{1-3} 2  &  4 & 6  \\
\hline
\end{tabular}
=\nonumber
\\
&\frac{2\sqrt{3}}{9}\vert I^1_2 \rangle \otimes \vert S^2_5 \rangle-\frac{\sqrt{6}}{9}\vert I^1_3 \rangle \otimes \vert S^2_5 \rangle+\frac{2\sqrt{3}}{9}\vert I^1_4 \rangle \otimes \vert S^2_3 \rangle- \nonumber\\ 
&\frac{\sqrt{6}}{9}\vert I^1_4 \rangle \otimes \vert S^2_4 \rangle-\frac{\sqrt{6}}{9}\vert I^1_5 \rangle \otimes \vert S^2_5 \rangle+\frac{\sqrt{3}}{9}\vert I^1_6 \rangle \otimes \vert S^2_5 \rangle- \nonumber\\ 
&\frac{\sqrt{6}}{9}\vert I^1_7 \rangle \otimes \vert S^2_3 \rangle+\frac{\sqrt{3}}{9}\vert I^1_7 \rangle \otimes \vert S^2_4 \rangle+\frac{\sqrt{5}}{5}\vert I^1_9 \rangle \otimes \vert S^2_1 \rangle- \nonumber\\
&\frac{\sqrt{270}}{45}\vert I^1_9 \rangle \otimes \vert S^2_2 \rangle.
\end{align}
We find the color $\otimes$ isospin $\otimes$ spin state satisfying the fully antisymmetry property for (I,S)=(1,2) to be given by, 
\begin{align}
\vert C, I^1, S^2 \rangle=\frac{1}{\sqrt{5}}(&\vert C_1 \rangle \otimes \vert [I^1, S^2]_5 \rangle-\vert C_2\rangle                           \otimes \vert [I^1, S^2]_4 \rangle-  \nonumber\\                                                                                                                     
&\vert C_3 \rangle \otimes \vert [I^1, S^2]_3 \rangle+\vert C_4 \rangle \otimes \vert [I^1, S^2]_2 \rangle-    \nonumber\\                                                                                                                     
&\vert C_5 \rangle \otimes \vert [I^1, S^2]_1 \rangle).
\end{align}

In the case of (I,S)=(2,1), the $\vert [I^2, S^1] \rangle$ basis
functions belonging to the Young tableau of [3,3] are presented as the followings :

\begin{align}
&\begin{tabular}{c}
$\vert [I^2, S^1]_1 \rangle$=
\end{tabular}
\begin{tabular}{|c|c|c|}
\hline
                1   & 2   & 3    \\
\cline{1-3} 4  &  5 & 6  \\
\hline
\end{tabular}
=\nonumber
\\
&\frac{\sqrt{15}}{9}\vert I^2_3 \rangle \otimes \vert S^1_1 \rangle+\frac{\sqrt{15}}{9}\vert I^2_2 \rangle \otimes \vert S^1_2 \rangle+\frac{2}{9}\vert I^2_3 \rangle \otimes \vert S^1_2 \rangle- \nonumber\\ 
&\frac{1}{9}\vert I^2_4 \rangle \otimes \vert S^1_3 \rangle-\frac{1}{9}\vert I^2_5 \rangle \otimes \vert S^1_4 \rangle+\frac{\sqrt{5}}{5}\vert I^2_1 \rangle \otimes \vert S^1_5 \rangle+ \nonumber\\ 
&\frac{\sqrt{120}}{45}\vert I^2_2 \rangle \otimes \vert S^1_5 \rangle+\frac{2\sqrt{2}}{9}\vert I^2_3 \rangle \otimes \vert S^1_5 \rangle-\frac{\sqrt{2}}{9}\vert I^2_4 \rangle \otimes \vert S^1_6 \rangle- \nonumber\\
&\frac{\sqrt{2}}{9}\vert I^2_5 \rangle \otimes \vert S^1_7 \rangle-\frac{\sqrt{6}}{9}\vert I^2_4 \rangle \otimes \vert S^1_8 \rangle-\frac{\sqrt{6}}{9}\vert I^2_5 \rangle \otimes \vert S^1_9 \rangle.
\end{align}
\begin{align}
&\begin{tabular}{c}
$\vert [I^2, S^1]_2 \rangle$=
\end{tabular}
\begin{tabular}{|c|c|c|}
\hline
                1   & 2   & 4    \\
\cline{1-3} 3  &  5 & 6  \\
\hline
\end{tabular}
=\nonumber
\\
&\frac{\sqrt{15}}{9}\vert I^2_4 \rangle \otimes \vert S^1_1 \rangle-\frac{1}{9}\vert I^2_4 \rangle \otimes \vert S^1_2 \rangle+\frac{\sqrt{15}}{9}\vert I^2_2 \rangle \otimes \vert S^1_3 \rangle- \nonumber\\ 
&\frac{1}{9}\vert I^2_3 \rangle \otimes \vert S^1_3 \rangle+\frac{\sqrt{2}}{9}\vert I^2_4 \rangle \otimes \vert S^1_3 \rangle-\frac{\sqrt{2}}{9}\vert I^2_5 \rangle \otimes \vert S^1_4 \rangle- \nonumber\\ 
&\frac{\sqrt{2}}{9}\vert I^2_4 \rangle \otimes \vert S^1_5 \rangle+\frac{\sqrt{5}}{5}\vert I^2_1 \rangle \otimes \vert S^1_6 \rangle+\frac{\sqrt{120}}{45}\vert I^2_2 \rangle \otimes \vert S^1_6 \rangle- \nonumber\\
&\frac{\sqrt{2}}{9}\vert I^2_3 \rangle \otimes \vert S^1_6 \rangle+\frac{2}{9}\vert I^2_4 \rangle \otimes \vert S^1_6 \rangle-\frac{2}{9}\vert I^2_5 \rangle \otimes \vert S^1_7 \rangle- \nonumber\\
&\frac{\sqrt{6}}{9}\vert I^2_3 \rangle \otimes \vert S^1_8 \rangle-\frac{\sqrt{3}}{9}\vert I^2_4 \rangle \otimes \vert S^1_8 \rangle+\frac{\sqrt{3}}{9}\vert I^2_5 \rangle \otimes \vert S^1_9 \rangle.
\end{align}
\begin{align}
&\begin{tabular}{c}
$\vert [I^2, S^1]_3 \rangle$=
\end{tabular}
\begin{tabular}{|c|c|c|}
\hline
                1   & 3   & 4    \\
\cline{1-3} 2  &  5 & 6  \\
\hline
\end{tabular}
=\nonumber
\\
&\frac{\sqrt{15}}{9}\vert I^2_5 \rangle \otimes \vert S^1_1 \rangle-\frac{1}{9}\vert I^2_5 \rangle \otimes \vert S^1_2 \rangle-\frac{\sqrt{2}}{9}\vert I^2_5 \rangle \otimes \vert S^1_3 \rangle+ \nonumber\\ 
&\frac{\sqrt{15}}{9}\vert I^2_2 \rangle \otimes \vert S^1_4 \rangle-\frac{1}{9}\vert I^2_3 \rangle \otimes \vert S^1_4 \rangle-\frac{\sqrt{2}}{9}\vert I^2_4 \rangle \otimes \vert S^1_4 \rangle- \nonumber\\ 
&\frac{\sqrt{2}}{9}\vert I^2_5 \rangle \otimes \vert S^1_5 \rangle-\frac{2}{9}\vert I^2_5 \rangle \otimes \vert S^1_6 \rangle+\frac{\sqrt{5}}{5}\vert I^2_1 \rangle \otimes \vert S^1_7 \rangle+ \nonumber\\
&\frac{\sqrt{120}}{45}\vert I^2_2 \rangle \otimes \vert S^1_7 \rangle-\frac{\sqrt{2}}{9}\vert I^2_3 \rangle \otimes \vert S^1_7 \rangle-\frac{2}{9}\vert I^2_4 \rangle \otimes \vert S^1_7 \rangle+ \nonumber\\
&\frac{\sqrt{3}}{9}\vert I^2_5 \rangle \otimes \vert S^1_8 \rangle-\frac{\sqrt{6}}{9}\vert I^2_3 \rangle \otimes \vert S^1_9 \rangle+\frac{\sqrt{3}}{9}\vert I^2_4 \rangle \otimes \vert S^1_9 \rangle.
\end{align}
\begin{align}
&\begin{tabular}{c}
$\vert [I^2, S^1]_4 \rangle$=
\end{tabular}
\begin{tabular}{|c|c|c|}
\hline
                1   & 2   & 5    \\
\cline{1-3} 3  &  4 & 6  \\
\hline
\end{tabular}
=\nonumber
\\
&\frac{2\sqrt{3}}{9}\vert I^2_4 \rangle \otimes \vert S^1_2 \rangle+\frac{2\sqrt{3}}{9}\vert I^2_3 \rangle \otimes \vert S^1_3 \rangle+\frac{\sqrt{6}}{9}\vert I^2_4 \rangle \otimes \vert S^1_3 \rangle- \nonumber\\ 
&\frac{\sqrt{6}}{9}\vert I^2_5 \rangle \otimes \vert S^1_4 \rangle-\frac{\sqrt{6}}{9}\vert I^2_4 \rangle \otimes \vert S^1_5 \rangle-\frac{\sqrt{6}}{9}\vert I^2_3 \rangle \otimes \vert S^1_6 \rangle- \nonumber\\ 
&\frac{\sqrt{3}}{9}\vert I^2_4 \rangle \otimes \vert S^1_6 \rangle+\frac{\sqrt{3}}{9}\vert I^2_5 \rangle \otimes \vert S^1_7 \rangle+\frac{\sqrt{5}}{5}\vert I^2_1 \rangle \otimes \vert S^1_8 \rangle- \nonumber\\
&\frac{\sqrt{270}}{45}\vert I^2_2 \rangle \otimes \vert S^1_8 \rangle.
\end{align}
\begin{align}
&\begin{tabular}{c}
$\vert [I^2, S^1]_5 \rangle$=
\end{tabular}
\begin{tabular}{|c|c|c|}
\hline
                1   & 3   & 5    \\
\cline{1-3} 2  &  4 & 6  \\
\hline
\end{tabular}
=\nonumber
\\
&\frac{2\sqrt{3}}{9}\vert I^2_5 \rangle \otimes \vert S^1_2 \rangle-\frac{\sqrt{6}}{9}\vert I^2_5 \rangle \otimes \vert S^1_3 \rangle+\frac{2\sqrt{3}}{9}\vert I^2_3 \rangle \otimes \vert S^1_4 \rangle- \nonumber\\ 
&\frac{\sqrt{6}}{9}\vert I^2_4 \rangle \otimes \vert S^1_4 \rangle-\frac{\sqrt{6}}{9}\vert I^2_5 \rangle \otimes \vert S^1_5 \rangle+\frac{\sqrt{3}}{9}\vert I^2_5 \rangle \otimes \vert S^1_6 \rangle- \nonumber\\ 
&\frac{\sqrt{6}}{9}\vert I^2_3 \rangle \otimes \vert S^1_7 \rangle+\frac{\sqrt{3}}{9}\vert I^2_4 \rangle \otimes \vert S^1_7 \rangle+\frac{\sqrt{5}}{5}\vert I^2_1 \rangle \otimes \vert S^1_9 \rangle- \nonumber\\
&\frac{\sqrt{270}}{45}\vert I^2_2 \rangle \otimes \vert S^1_9 \rangle.
\end{align}
We find the color $\otimes$ isospin $\otimes$ spin state satisfying the fully antisymmetry property for (I,S)=(2,1) to be given by, 
\begin{align}
\vert C, I^2, S^1 \rangle=\frac{1}{\sqrt{5}}(&\vert C_1 \rangle \otimes \vert [I^2, S^1]_5 \rangle-\vert C_2\rangle \otimes \vert [I^2, S^1]_4 \rangle-  \nonumber\\                                                                                                                     
&\vert C_3 \rangle \otimes \vert [I^2, S^1]_3 \rangle+\vert C_4 \rangle \otimes \vert [I^2, S^1]_2 \rangle-    \nonumber\\                                                                                                                     
&\vert C_5 \rangle \otimes \vert [I^2, S^1]_1 \rangle).
\end{align}

For the case of (I,S)=(3,0) and (I,S)=(0,3), we find straightforwardly the fully antisymmetric color $\otimes$ isospin $\otimes$ spin state, written by, respectively, 
\begin{align}
\vert C, I^3, S^0 \rangle=\frac{1}{\sqrt{5}}(&\vert C_1 \rangle \otimes \vert I^3 \rangle \otimes \vert S^0_5 \rangle-\vert C_2\rangle \otimes \vert I^3 \rangle \otimes \vert S^0_4 \rangle-  \nonumber\\                                                                                                                     
&\vert C_3 \rangle \otimes \vert I^3 \rangle \otimes \vert S^0_3 \rangle+\vert C_4 \rangle \otimes \vert I^3 \rangle \otimes \vert S^0_2 \rangle-    \nonumber\\                                                                                                                     
&\vert C_5 \rangle \otimes \vert I^3 \rangle \otimes \vert S^0_1 \rangle),
\end{align}
\begin{align}
\vert C, I^0, S^3 \rangle=\frac{1}{\sqrt{5}}(&\vert C_1 \rangle \otimes \vert I^0_5 \rangle \otimes \vert S^3 \rangle-\vert C_2\rangle \otimes \vert I^0_4 \rangle \otimes \vert S^3 \rangle-  \nonumber\\                                                                                                                     
&\vert C_3 \rangle \otimes \vert I^0_3 \rangle \otimes \vert S^3 \rangle+\vert C_4 \rangle \otimes \vert I^0_2 \rangle \otimes \vert S^3 \rangle-    \nonumber\\                                                                                                                     
&\vert C_5 \rangle \otimes \vert I^0_1 \rangle \otimes \vert S^3 \rangle).
\end{align}

In dealing with the expectation value of $-\lambda_i^c\lambda_j^c{\sigma}_i\cdot{\sigma}_j$ with respect
to $\vert C, I, S \rangle$ state for all of (I,S), the symmetry property of the state make this calculation simple in that $\langle -{\sum}_{i<j}^{6}\lambda_i^c\lambda_j^c{\sigma}_i\cdot{\sigma}_j \rangle$=15$\langle -\lambda_1^c\lambda_2^c{\sigma}_1\cdot{\sigma}_2 \rangle$ as argued early.  Moreover, using the symmetry properties of the wave function, one can derive the effective 
formula which is expressed in terms of the Casimir operators of the isospin, spin, and color, given by~\cite{Aerts:1977rw},
\begin{align}
-&{\sum}_{i<j}^{N}\lambda_i^c\lambda_j^c{\sigma}_i\cdot{\sigma}_j = \nonumber\\                                                                                                                     
[&\frac{4}{3}N(N-6)+4I(I+1)+\frac{4}{3}S(S+1)+2C_c],
\end{align}
where N is the total number of quarks in this system, and $C_c$=$\frac{1}{4}\lambda^c\lambda^c$, that is,
the first kind of Casimir operator of SU(3) in the system of N quarks. Since we must consider only  the color singlet as
a physical observable, the term $C_c$ vanishes. For practical purposes, our the  calculation of $\langle -\lambda_1^c\lambda_2^c{\sigma}_1\cdot{\sigma}_2 \rangle$ can be easily performed with the symmetry of between particles 1 and 2 in the $\vert C, I, S \rangle$ state, whose the property of the symmetry is definitely derived from the
Young tableau. Then we find  the following formula: 
\begin{align}
&\begin{tabular}{c}
$\lambda_1^c\lambda_2^c$ $\vert$ 
\end{tabular}
\begin{tabular}{|c|}
\hline
1   \\
\hline
2   \\
\hline
\end{tabular} 
\rangle=-\frac{8}{3}\vert
\begin{tabular}{|c|}
\hline
1   \\
\hline
2   \\
\hline
\end{tabular}
\rangle,
\begin{tabular}{c}
$\lambda_1^c\lambda_2^c$ $\vert$ 
\end{tabular}
\begin{tabular}{|c|c|}
\hline
1  & 2 \\
\hline
\end{tabular} 
\rangle=\frac{4}{3}\vert
\begin{tabular}{|c|c|}
\hline
1  & 2 \\
\hline
\end{tabular}
\rangle, \nonumber \\
&\begin{tabular}{c}
${\sigma}_1\cdot{\sigma}_2$ $\vert$ 
\end{tabular}
\begin{tabular}{|c|}
\hline
1   \\
\hline
2   \\
\hline
\end{tabular} 
\rangle=-\vert
\begin{tabular}{|c|}
\hline
1   \\
\hline
2   \\
\hline
\end{tabular}
\rangle,
\begin{tabular}{c}
${\sigma}_1\cdot{\sigma}_2$ $\vert$ 
\end{tabular}
\begin{tabular}{|c|c|}
\hline
1  & 2 \\
\hline
\end{tabular} 
\rangle=\vert
\begin{tabular}{|c|c|}
\hline
1  & 2 \\
\hline
\end{tabular}
\rangle.
\end{align}
\begin{table}[htdp]
\caption{The expectation value of $-{\sum}_{i<j}^{6}\lambda_i^c\lambda_j^c{\sigma}_i\cdot{\sigma}_j$  with respect to  $\vert C, I, S \rangle$  for  all possible (I,S) configurations  in the dibaryon, which will be denoted by $V_d$. $\Delta V$ is $V_d$-($V_{b1}$+$V_{b2}$), in which the $V_{b1}$ and $V_{b2}$ are those of the baryons to which the dibaryon can decay.  }
\begin{center}
\begin{tabular}{c|c|c|c|c|c|c}
\hline \hline
    (I,S)      &  (3,0)     &  (2,1)    & (1,2)    & (1,0)   &  (0,3)     & (0,1)  \\
\hline
$V_d$ &48   & $\frac{80}{3}$ & 16  &8       &     16     & $\frac{8}{3}$         \\
\hline 
$\Delta V$  &32  & $\frac{80}{3}$ & 16  & 24       &     0     & $\frac{56}{3}$         \\
\hline  \hline
\end{tabular}
\end{center}
\label{normalmeson_mass-0}
\end{table}

\section{Numerical results}

\begin{table}[htdp]
\caption{The mass of the dibaryon in (I,S) state with the two types of potentials given in Eq.~(2) and Eq.~(3). The binding energy $E_B$ is taken to be the difference between the mass of the dibaryon and the two baryon threshold. The dimension of the variational parameters are given in fm$^{-2}$. }
\begin{center}
\begin{tabular}{c|c|c|c|c|c|c}
\hline \hline
    (I,S)                &  (3,0)          &  (2,1)         & (1,2)         & (1,0)        &  (0,3)         & (0,1)  \\
\hline
 Type 1               &   3132.3      &  2926.5     &  2808.5    &  2710.8    &   2808.5     & 2639.6          \\
\hline
Variational           & $a$=1.2,        &  $a$=0.9,     & $a$=1.4,       & $a$=3,        &  $a$=1.4,        &  $a$=3.6,          \\
parameters          &  $b$=1.2,       & $b$=1.5,      & $b$=1.8,       & $b$=1.4,    &  $b$=1.8,         &  $b$=1.4,         \\
                          &  $c$=1.4        & $c$=1.7       & $c$=0.9        &  $c$=1.3     &  $c$=0.9         &   $c$=1.3          \\
\hline
 $E_B$             & 597.3        & 681.9      & 563.9        & 756.6       & 273.5       & 685.4                   \\
\hline \hline
 Type 2                  &  2845.5      &  2711.7    &  2629.9     & 2558.6      & 2629.9      &     2504.1     \\
\hline
Variational              &  $a$=0.6,      &  $a$=0.5,     &  $a$=0.7,       & $a$=2.2,      &  $a$=0.7,      &     $a$=2.6,      \\
parameters             &  $b$=0.7,      &  $b$=0.9,     &  $b$=1.1,      & $b$=0.8,      &  $b$=1.1,      &     $b$=0.8,      \\
                             &  $c$=0.6       &  $c$=0.8       &  $c$=0.5       & $c$=0.7        & $c$=0.5        &     $c$=0.8       \\
\hline
 $E_B$              & 371.1    & 498.2      & 416.4      & 606.0      & 155.5    & 551.5         \\
\hline \hline
\end{tabular}
\end{center}
\label{normalmeson_mass1}
\end{table}

In this section, we analyze the numerical results obtained from the variational method, by using the completely symmetric spatial function as the trial function.
Table \ref{normalmeson_mass1} shows the result of the analysis with the trial spacial wave function given in Eq.~(\ref{eq-jac1}) after adding 14 additional forms as discussed before.   
 Among all the dibaryons with (I,S), it is the dibaryon with (I=0,S=3) that is most likely to be stable against the strong decay. However, as we see in the Table \ref{normalmeson_mass1} , we find that even for this state the two baryon threshold lies below the dibaryon mass. 
 
 Although simple comparison of $\langle -{\sum}_{i<j}^{6}\lambda_i^c\lambda_j^c{\sigma}_i\cdot{\sigma}_j \rangle$ in the hyperfine interaction between the dibaryon and two baryons, as given in Table.~\ref{normalmeson_mass-0}, shows the splitting is minimal in the (I=0,S=3) channel, the lowest mass of the dibaryon for both types of the potential considered is far above the threshold of two $\Delta$ baryons. 
 
Table~\ref{normalmeson_mass1} shows that the dibaryon mass is larger for type 1 potential compared to that for type 2.  This is because the  size of the dibaryon in the former is smaller than that in the latter, as determined by the inverse of the variational parameters $a,b,c$. To better understand this point, we show the values of each energy term of the dibaryon with (I=0,S=3) in Table~\ref{normalmeson_mass2}.
\begin{table}[htdp]
\caption{The  values of each energy term of the dibaryon with (I=0,S=3) and $
\Delta$ baryon. $\Delta$E is the difference between the dibaryon and two $
\Delta$ baryon in each term.  }
\begin{center}
\begin{tabular}{c|c|c|c|c}
\hline \hline
   Type 1                      & kinetic     & linear               &  coulomb         &  hyperfine        \\
\hline
 Dibaryon                   & 1282.8     &  2603.6              &  -446.9          &   186.9             \\
\hline
$\Delta$ Baryon         &   589.2     &  1214.8              & -239.3          &    111.6               \\
\cline{2-5}
 Variational parameters  &\multicolumn{4}{|c}{$a$=1.4\qquad $b$=2.1}                                            \\                                                   
 \hline
 $\Delta$E                 & 104.2     & 173.8                 & 31.6          &  -36.3             \\
\hline \hline
  Type 2                      & kinetic     & 1/2 power             &  coulomb         &  hyperfine        \\
\hline
 Dibaryon                   & 722.7         & 2924.1                 & -346.6             & 132.5               \\
\hline
$\Delta$ Baryon         & 333.2       & 1410.4                  & -186.6          &    81.7                \\
\cline{2-5}
 Variational parameters  &\multicolumn{4}{|c}{$a$=1.4\qquad $b$=0.7}                                            \\                                                   
 \hline
 $\Delta$E                 & 56.2        &  103.2                  & 26.7           &  -30.8              \\
\hline \hline
\end{tabular}
\end{center}
\label{normalmeson_mass2}
\end{table}

Since the size of the dibaryon in both types of potentials are larger than that of a single  baryon, the value of the kinetic part of the dibaryon is comparatively larger than the sum of that of the two $\Delta$ baryons.  Moreover, the value of kinetic part of dibaryon in type 2 is a little larger compared to that in type 1 due to its relatively small increase in size.

In addition to the kinetic term,  the effect of 1/2-power confinement part of the type 2 potential causes the mass of the dibaryon to decrease compared to the case of type 1.  Nevertheless, the lowest mass in type 2 is still about 155 $\rm{MeV}$ above the threshold of two $\Delta$ baryons, and no other choice for the confinement potential, such as 1/3-power, is expected to change the stability of the dibaryon. 

In investigating the stability in the present work, we can consider three-body color confinement operators which are mentioned in the introduction, as this approach may change the stability of multiquark configurations, such as that of the dibaryon. The two types of  operators 
can be expressed in terms of the permutation operators given by
\begin{align}
d^{abc}F^a_i F^b_j F^c_k=&\frac{1}{4}[(ijk)+(ikj)]+\frac{1}{9}I \nonumber \\
&-\frac{1}{6}[(ij)+(ik)+(jk)], \nonumber 
\end{align} 
\begin{align}
f^{abc}F^a_i F^b_j F^c_k=-\frac{i}{4}[(ijk)-(ikj)],
\end{align} 
where I is the identity operator, and $(ijk)$ and $(ij)$ are operators belonging to the permutation group of $S_6$, which is called 3-cycles, and 2-cycles, respectively, and $F^a_i$=$1/2{\lambda}^a_i$. Also, there is
another formula for $d^{abc}F^a_i F^b_j F^c_k$, which can be conveniently shown to be invariant under the SU(3) algebra, given by~\cite{Pepin:2001is},
\begin{align}
d^{abc}F^a_i F^b_j F^c_k=\frac{1}{6}[C^{(3)}_{i+j+k}-\frac{5}{2}C^{(2)}_{i+j+k}+\frac{20}{3}],
\end{align}
where $C^{(2)}$ is the first kind of Casimir operator, and $C^{(2)}$, the second kind of Casimir operator of SU(3). Since baryon consists of three quarks, the SU(3) invariant operators are written by,
\begin{align}
d^{abc}F^a_1 F^b_2 F^c_3=&\frac{1}{4}[(123)+(132)]+\frac{1}{9}I \nonumber \\
&-\frac{1}{6}[(12)+(13)+(23)], \nonumber 
\end{align}
\begin{align}
f^{abc}F^a_1 F^b_2 F^c_3=-\frac{i}{4}[(123)-(132)].
\end{align} 
For baryon which has one color singlet represented by the standard Young-Yamanouchi basis of Young tableau [1,1,1], we use the irreducible representation with one dimension given by                 
\begin{align}
\begin{tabular}{c}
$\sigma$  
\end{tabular}
\begin{tabular}{|c|}
\hline
1   \\
\hline
2   \\
\hline
3    \\
\hline
\end{tabular} 
=(-1)^{\sigma}
\begin{tabular}{|c|}
\hline
1   \\
\hline
2   \\
\hline
3    \\
\hline
\end{tabular} 
, \qquad (\sigma \in S_3)
\end{align}
where $(-1)^{\sigma}$ is 1 if $\sigma$ is an even permutation, and -1 if $\sigma$ is an odd permutation. By using this formula, we can easily calculate the action of the operators on the color singlet of baryon, $\vert C \rangle$=$\frac{1}{\sqrt{6}}\epsilon_{ijk}q^i(1)q^j(2)q^k(3))$ in Eq.~(10) ;
\begin{align}
&d^{abc}F^a_1 F^b_2 F^c_3 \vert C \rangle =\frac{10}{9} \vert C \rangle , \label{d-type-b} \\
&f^{abc}F^a_1 F^b_2 F^c_3 \vert C \rangle =0. \label{f-type-b}
\end{align} 
As we can see, while the introduction of $d^{abc}F^a_1 F^b_2 F^c_3$ can contribute to the  baryon mass, $f^{abc}F^a_1 F^b_2 F^c_3$ will not.

Likewise, for dibaryon which has five color singlet bases corresponding to the standard Young-Yamanuochi bases of Young tableau [2,2,2], we can calculate the irreducible matrix representation of the permutation operators belonging to $S_6$ in the form of 5$\times$5 matrix in terms of the five color singlet bases. Since the irreducible matrix of (ij) and [(ijk) + (ikj)] are symmetric, and the irreducible matrix of (ijk) is antisymmetric, $d^{abc}F^a_i F^b_j F^c_k$ has a form of symmetric matrix and $f^{abc}F^a_i F^b_j F^c_k$ has a form of hermitian matrix with vanishing diagonal elements. As we will show in detail in appendix, the matrix of $d^{abc}F^a_1 F^b_2 F^c_3$ and $f^{abc}F^a_1 F^b_2 F^c_3$ are given by,
\begin{align}
&\langle C_1 \vert d^{abc}F^a_1 F^b_2 F^c_3 \vert C_1 \rangle= \langle C_2 \vert d^{abc}F^a_1 F^b_2 F^c_3 \vert C_2 \rangle= \nonumber \\
&\langle C_3 \vert d^{abc}F^a_1 F^b_2 F^c_3 \vert C_3 \rangle=\langle C_4 \vert d^{abc}F^a_1 F^b_2 F^c_3 \vert C_4 \rangle=-\frac{5}{36} \nonumber \\
&\langle C_5 \vert d^{abc}F^a_1 F^b_2 F^c_3 \vert C_5 \rangle=10/9, \nonumber \\
&\langle C_1 \vert f^{abc}F^a_1 F^b_2 F^c_3 \vert C_1 \rangle= \langle C_2 \vert f^{abc}F^a_1 F^b_2 F^c_3 \vert C_2 \rangle= \nonumber \\
&\langle C_3 \vert f^{abc}F^a_1 F^b_2 F^c_3 \vert C_3 \rangle=\langle C_4 \vert f^{abc}F^a_1 F^b_2 F^c_3 \vert C_4 \rangle= \nonumber \\
&\langle C_5 \vert f^{abc}F^a_1 F^b_2 F^c_3 \vert C_5 \rangle=0.
\end{align}
Then, we can find the expectation value of the three-body confinement operators, $d^{abc}F^a_1 F^b_2 F^c_3$, and $f^{abc}F^a_1 F^b_2 F^c_3$ in terms of $\vert C, I^i, S^j \rangle$ for (I=i,S=j), reminding that
\begin{align}
\vert C, I^i, S^j \rangle=\frac{1}{\sqrt{5}}(&\vert C_1 \rangle \otimes \vert [I^i, S^j]_5 \rangle-\vert C_2\rangle \otimes \vert [I^i, S^j]_4 \rangle-  \nonumber\\                                                                                                                     
&\vert C_3 \rangle \otimes \vert [I^i, S^j]_3 \rangle+\vert C_4 \rangle \otimes \vert [I^i, S^j]_2 \rangle-    \nonumber\\                                                                                                                     
&\vert C_5 \rangle \otimes \vert [I^i, S^j]_1 \rangle),
\end{align}
one finds the followings:
\begin{align}
\langle C, I^i, S^j \vert d^{abc}F^a_1 F^b_2 F^c_3 \vert C, I^i, S^j \rangle&=\frac{1}{5}(-4\times\frac{5}{36}+\frac{10}{9}) \nonumber \\
&=\frac{1}{9}, \label{d-type} 
\end{align}
\begin{align}
&\langle C, I^i, S^j \vert f^{abc}F^a_1 F^b_2 F^c_3 \vert C, I^i, S^j \rangle=0.
\label{f-type}
\end{align}
Due to the complete symmetry of $\vert$ $C, I^i, S^j$ $\rangle$ under any permutation of $S_6$,  one finds that     
 $\langle$ $C, I^i, S^j$ $\vert$ $d^{abc}F^a_l F^b_m F^c_n$ $\vert$ $C, I^i, S^j$ $\rangle$ =$\langle$ $C, I^i, S^j$ $\vert$ $d^{abc}F^a_1 F^b_2 F^c_3$ $\vert$ $C, I^i, S^j$ $\rangle$ 
for any $l, m, n$ ($l$$<$$m$$<$$n$=1$\sim$6).  Similar relation holds for  $\langle$ $C, I^i, S^j$ $\vert$ $f^{abc}F^a_l F^b_m F^c_n$ $\vert$ $C, I^i, S^j$ $\rangle$.
As far as the color factors are concerned, one finds that the contribution of the $d$-type three body interaction contributing to the dibaryon as given by  $C^3_6$ = 20
times Eq.~(\ref{d-type}) is equal to twice that of the baryon given in Eq.~(\ref{d-type-b}).

 If we attempt to add the $d$-type three-body confinement potential, the result will depend on the detailed functional form multiplying the SU(3) invariant operators.
For the simplest choice, we can chose it to be the sum of two body potential such as 
( $V_{123}$ = $d^{abc}F^a_1 F^b_2 F^c_3$ ( $r_{12}$/$a_0$ +$r_{13}$/$a_0$ + $r_{23}$/$a_0$) ).  However, for such a simplified form, the result will only be a re-parametrization of the parameters of the model and there will be no change in the stability of the dibaryon.  
On the other hand, a intrinsic three body force will change the situation and on that grounds, it is of great importance to have some ideas on the understanding of the  two-body and three-body confinement.

\section{Summary}

In order for the total wave function to be fully antisymmetric in six quarks system with only $u$, $d$ quarks, we first  
consider the spatial function that is fully symmetric, and find 15 Jacobian coordinates appropriate for the symmetry, and then, construct the spatial function desirable for this scheme with Gaussian spatial
function to perform variational method in nonrelativitic Hamiltonian. Secondly, we classify the physical states with respect to isospin (I) and spin (S), and find the color singlet basis functions, isospin basis functions, and spin basis functions allowed to the six quarks system, and construct the color $\otimes$
isospin $\otimes$ spin states that should be completely antisymmetric, by means of IS scheme that couples the color basis function to IS basis function.  
We find that there does not exist a compact dibayron system in all system that is stable against the decay into two bayrons with corresponding quantum numbers. 
Hence, the recently observed peak in the $I=0,S=3$ B=2 channel~\cite{Bashkanov:2008ih,Adlarson:2011bh,Adlarson:2012fe,Adlarson:2014pxj,Adlarson:2014ozl,Adlarson:2014tcn} should be a molecular configuration composed of two $\Delta$ dibaryons.

\appendix

\section{}

In this section, we derive the three-body color operators, which is invariant to SU(3), in terms of the relevant permutation operators. This will enable us to represent the the three-body color operator with respect to color singlet basis functions of the dibaryon. We can express the algebra of SU(3) as the permutation of two particles, as in the case for the algebra of SU(2) ; 
\begin{align}
&(12)=\frac{1}{2}I+\frac{1}{2}{\sigma}_1\cdot{\sigma}_2, \nonumber \\
&(12)=\frac{1}{3}I+\frac{1}{2}{\lambda}^c_1{\lambda}^c_2.
\end{align}
Here, I is the identity operator, and (12) is 2-cycles permutation. Then, 
noting that (123), and (132) can be written as (123) = (23) $\cdot$ (12) and (132) = (23) $\cdot$ (13), we can straightforwardly present 3-cycles permutation as~\cite{Dmitrasinovic:2001nv};
\begin{align}
&(123)=(23)\cdot(12)=(\frac{1}{3}I+\frac{1}{2}{\lambda}^c_2{\lambda}^c_3)(\frac{1}{3}I+\frac{1}{2}{\lambda}^c_1{\lambda}^c_2)= \nonumber \\
&\frac{1}{9}I+\frac{1}{6}\sum_{i<j}^{3}\lambda^c_i\lambda^c_j+2d^{abc}F^a_1 F^b_2 F^c_3+
2if^{abc}F^a_1 F^b_2 F^c_3.
\end{align}
\begin{align}
&(132)=(23)\cdot(13)=(\frac{1}{3}I+\frac{1}{2}{\lambda}^c_2{\lambda}^c_3)(\frac{1}{3}I+\frac{1}{2}{\lambda}^c_1{\lambda}^c_3)= \nonumber \\
&\frac{1}{9}I+\frac{1}{6}\sum_{i<j}^{3}\lambda^c_i\lambda^c_j+2d^{abc}F^a_1 F^b_2 F^c_3-
2if^{abc}F^a_1 F^b_2 F^c_3.
\end{align}
By adding Eq.~(A3) to Eq.~(A2) and subtracting Eq.~(A3) from Eq.~(A2), we obtain $d^{abc}F^a_1 F^b_2 F^c_3$ and $f^{abc}F^a_1 F^b_2 F^c_3$. Also, we can apply this process to any (ijk) (i$<$j$<$k=1$\sim$6), and finally obtain the Eq.~(52) ;
\begin{align}
d^{abc}F^a_i F^b_j F^c_k=&\frac{1}{4}[(ijk)+(ikj)]+\frac{1}{9}I \nonumber \\
&-\frac{1}{6}[(ij)+(ik)+(jk)], \nonumber 
\end{align}
\begin{align}
f^{abc}F^a_i F^b_j F^c_k=-\frac{i}{4}[(ijk)-(ikj)].
\end{align} 

Now, we can construct the matrix representation of the three-body color operators in terms of the standard Young-Yamanouchi bases corresponding to the color singlet bases of dibaryon. As mentioned earlier, the standard Young-Yamanouchi bases which is orthonormal to each other are written by, 
\begin{align}
&\begin{tabular}{c}
$\vert C_1 \rangle$=
\end{tabular}
\begin{tabular}{|c|c|}
\hline
1 & 2   \\
\cline{1-2}
3 &  4  \\
\cline{1-2}
5 &  6  \\
\hline
\end{tabular} 
,
\begin{tabular}{c}
$\vert C_2 \rangle$=
\end{tabular}
\begin{tabular}{|c|c|}
\hline
1 & 3   \\
\cline{1-2}
2 &  4  \\
\cline{1-2}
5 &  6  \\
\hline
\end{tabular} 
,
\begin{tabular}{c}
$\vert C_3 \rangle$=
\end{tabular}
\begin{tabular}{|c|c|}
\hline
1 & 2   \\
\cline{1-2}
3 &  5  \\
\cline{1-2}
4 &  6  \\
\hline
\end{tabular}
,
\begin{tabular}{c}
$\vert C_4 \rangle$=
\end{tabular}
\begin{tabular}{|c|c|}
\hline
1 & 3   \\
\cline{1-2}
2 &  5  \\
\cline{1-2}
4 &  6  \\
\hline
\end{tabular}
,
\begin{tabular}{c}
$\vert C_5 \rangle$=
\end{tabular}
\begin{tabular}{|c|c|}
\hline
1 & 4   \\
\cline{1-2}
2 &  5  \\
\cline{1-2}
3 &  6  \\
\hline
\end{tabular} 
.
\end{align}
\begin{widetext}\allowdisplaybreaks
The matrix of $d^{abc}F^a_i F^b_j F^c_k$ is given by, 
\begin{align}
d_{abc}F^a_1F^b_2F^c_3=\left(\begin{array}{ccccc}
-\frac{5}{36}       &    0     & 0           & 0                                 & 0                                       \\  
0                  &   -\frac{5}{36}            & 0              & 0                             & 0                   \\  
0                  &            0              & -\frac{5}{36}       &      0            & 0                          \\ 
0                  &             0         &  0                        & -\frac{5}{36}             & 0               \\
0                  &       0             &     0                 &       0         &   \frac{10}{9}                  \\                                                      
\end{array} \right),\nonumber 
d_{abc}F^a_1F^b_2F^c_4=\left(\begin{array}{ccccc}
-\frac{5}{36}  &    0             & 0                      & 0                    & 0                              \\  
0                  &   -\frac{5}{36}         & 0              & 0                             & 0                   \\  
0                  &            0              & -\frac{5}{36}    &      0                   & 0                    \\ 
0                  &             0         &  0                       &  \frac{35}{36}    & \frac{5}{9\sqrt{2}} \\
0                  &       0             &     0                 &   \frac{5}{9\sqrt{2}}     &   0               \\                                                      
\end{array} \right),
\end{align}
\begin{align}
d_{abc}F^a_1F^b_2F^c_5=\left(\begin{array}{ccccc}
-\frac{5}{36}       &    0     & 0           & 0                                 & 0                                       \\  
0                  &  \frac{25}{36}    & 0           &  \frac{5}{6\sqrt{3}}  & -\frac{5}{6\sqrt{6}}        \\  
0                  &            0              & -\frac{5}{36}       &      0            & 0                          \\ 
0                  & \frac{5}{6\sqrt{3}} &  0         & \frac{5}{36}     & -\frac{5}{18\sqrt{2}}        \\
0                  & -\frac{5}{6\sqrt{6}}  &     0   & -\frac{5}{18\sqrt{2}}      &  0                       \\                                                      
\end{array} \right),\nonumber 
d_{abc}F^a_1F^b_2F^c_6=\left(\begin{array}{ccccc}
-\frac{5}{36}       &    0     & 0           & 0                                 & 0                                       \\  
0                  &  \frac{25}{36}    & 0           &  -\frac{5}{6\sqrt{3}}  & \frac{5}{6\sqrt{6}}        \\  
0                  &            0              & -\frac{5}{36}       &      0            & 0                          \\ 
0                  & -\frac{5}{6\sqrt{3}} &  0         & \frac{5}{36}     & -\frac{5}{18\sqrt{2}}        \\
0                  & \frac{5}{6\sqrt{6}}  &     0   & -\frac{5}{18\sqrt{2}}      &  0                       \\                                                      
\end{array} \right),
\end{align}
\begin{align}
d_{abc}F^a_1F^b_3F^c_4=\left(\begin{array}{ccccc}
-\frac{5}{36}       &    0     & 0           & 0                                 & 0                                       \\  
0                  & -\frac{5}{36}    & 0           &  0                    &      0                                    \\  
0                  &            0          & \frac{25}{36}   &\frac{5}{6\sqrt{3}}  & -\frac{5}{6\sqrt{6}}    \\ 
0                  & 0               & \frac{5}{6\sqrt{3}} & \frac{5}{36}     & -\frac{5}{18\sqrt{2}}        \\
0                  & 0               & -\frac{5}{6\sqrt{6}}   & -\frac{5}{18\sqrt{2}}      &  0                    \\                                                      
\end{array} \right),\nonumber 
d_{abc}F^a_1F^b_3F^c_5=\left(\begin{array}{ccccc}
\frac{35}{72} &\frac{5}{8\sqrt{3}}&\frac{5}{8\sqrt{3}}&\frac{5}{24}& \frac{5}{12\sqrt{2}}      \\                       
\frac{5}{8\sqrt{3}} &\frac{5}{72} & \frac{5}{24} &\frac{5}{24\sqrt{3}} &\frac{5}{12\sqrt{6}} \\  
\frac{5}{8\sqrt{3}} &\frac{5}{24} & \frac{5}{72} &\frac{5}{24\sqrt{3}} &\frac{5}{12\sqrt{6}} \\  
\frac{5}{24} &\frac{5}{24\sqrt{3}} & \frac{5}{24\sqrt{3}}&-\frac{5}{72}&\frac{5}{36\sqrt{2}} \\  
\frac{5}{12\sqrt{2}} &\frac{5}{12\sqrt{6}} & \frac{5}{12\sqrt{6}}&\frac{5}{36\sqrt{2}}&0 \\  
\end{array} \right),
\end{align}
\begin{align}
d_{abc}F^a_1F^b_3F^c_6=\left(\begin{array}{ccccc}
\frac{35}{72} &\frac{5}{8\sqrt{3}}&-\frac{5}{8\sqrt{3}}&-\frac{5}{24}& -\frac{5}{12\sqrt{2}}    \\                       
\frac{5}{8\sqrt{3}} &\frac{5}{72} & -\frac{5}{24} &-\frac{5}{24\sqrt{3}} &-\frac{5}{12\sqrt{6}} \\  
-\frac{5}{8\sqrt{3}} &-\frac{5}{24} & \frac{5}{72} &\frac{5}{24\sqrt{3}} &\frac{5}{12\sqrt{6}} \\  
-\frac{5}{24} &-\frac{5}{24\sqrt{3}} & \frac{5}{24\sqrt{3}}&-\frac{5}{72}&\frac{5}{36\sqrt{2}} \\  
-\frac{5}{12\sqrt{2}} &-\frac{5}{12\sqrt{6}} & \frac{5}{12\sqrt{6}}&\frac{5}{36\sqrt{2}}&0 \\  
\end{array} \right),\nonumber 
d_{abc}F^a_1F^b_4F^c_5=\left(\begin{array}{ccccc}
\frac{35}{72} &-\frac{5}{8\sqrt{3}}&-\frac{5}{8\sqrt{3}}&\frac{5}{24}& \frac{5}{12\sqrt{2}}    \\                       
-\frac{5}{8\sqrt{3}} &\frac{5}{72} & \frac{5}{24} &-\frac{5}{24\sqrt{3}} &-\frac{5}{12\sqrt{6}} \\  
-\frac{5}{8\sqrt{3}} &\frac{5}{24} & \frac{5}{72} &-\frac{5}{24\sqrt{3}} &-\frac{5}{12\sqrt{6}} \\  
\frac{5}{24} &-\frac{5}{24\sqrt{3}} & -\frac{5}{24\sqrt{3}}&-\frac{5}{72}&\frac{5}{36\sqrt{2}} \\  
\frac{5}{12\sqrt{2}} &-\frac{5}{12\sqrt{6}} & -\frac{5}{12\sqrt{6}}&\frac{5}{36\sqrt{2}}&0 \\  
\end{array} \right),
\end{align}
\begin{align}
d_{abc}F^a_1F^b_4F^c_6=\left(\begin{array}{ccccc}
\frac{35}{72} &-\frac{5}{8\sqrt{3}}&\frac{5}{8\sqrt{3}}&-\frac{5}{24}& -\frac{5}{12\sqrt{2}}    \\                       
-\frac{5}{8\sqrt{3}} &\frac{5}{72} & -\frac{5}{24} &\frac{5}{24\sqrt{3}} &\frac{5}{12\sqrt{6}} \\  
\frac{5}{8\sqrt{3}} &-\frac{5}{24} & \frac{5}{72} &-\frac{5}{24\sqrt{3}} &-\frac{5}{12\sqrt{6}} \\  
-\frac{5}{24} &\frac{5}{24\sqrt{3}} & -\frac{5}{24\sqrt{3}}&-\frac{5}{72}&\frac{5}{36\sqrt{2}} \\  
-\frac{5}{12\sqrt{2}} &\frac{5}{12\sqrt{6}} & -\frac{5}{12\sqrt{6}}&\frac{5}{36\sqrt{2}}&0 \\  
\end{array} \right),\nonumber 
d_{abc}F^a_1F^b_5F^c_6=\left(\begin{array}{ccccc}
-\frac{5}{36}       &    0     & 0           & 0                                 & 0                                       \\  
0                  & -\frac{5}{36}    & 0           &  0                    &      0                                    \\  
0                  &            0          & \frac{25}{36}   &-\frac{5}{6\sqrt{3}}  & \frac{5}{6\sqrt{6}}    \\ 
0                  & 0               & -\frac{5}{6\sqrt{3}} & \frac{5}{36}     & -\frac{5}{18\sqrt{2}}        \\
0                  & 0               & \frac{5}{6\sqrt{6}}   & -\frac{5}{18\sqrt{2}}      &  0                    \\                                                      
\end{array} \right),\nonumber
\end{align}
\begin{align}
d_{abc}F^a_2F^b_3F^c_4=\left(\begin{array}{ccccc}
-\frac{5}{36}       &    0     & 0           & 0                                 & 0                                       \\  
0                  & -\frac{5}{36}    & 0           &  0                    &      0                                    \\  
0                  &            0          & \frac{25}{36}   &-\frac{5}{6\sqrt{3}}  & \frac{5}{6\sqrt{6}}    \\ 
0                  & 0               & -\frac{5}{6\sqrt{3}} & \frac{5}{36}     & -\frac{5}{18\sqrt{2}}        \\
0                  & 0               & \frac{5}{6\sqrt{6}}   & -\frac{5}{18\sqrt{2}}      &  0                    \\                                                      
\end{array} \right),
d_{abc}F^a_2F^b_3F^c_5=\left(\begin{array}{ccccc}
\frac{35}{72} &-\frac{5}{8\sqrt{3}}&\frac{5}{8\sqrt{3}}&-\frac{5}{24}& -\frac{5}{12\sqrt{2}}    \\                       
-\frac{5}{8\sqrt{3}} &\frac{5}{72} & -\frac{5}{24} &\frac{5}{24\sqrt{3}} &\frac{5}{12\sqrt{6}} \\  
\frac{5}{8\sqrt{3}} &-\frac{5}{24} & \frac{5}{72} &-\frac{5}{24\sqrt{3}} &-\frac{5}{12\sqrt{6}} \\  
-\frac{5}{24} &\frac{5}{24\sqrt{3}} & -\frac{5}{24\sqrt{3}}&-\frac{5}{72}&\frac{5}{36\sqrt{2}} \\  
-\frac{5}{12\sqrt{2}} &\frac{5}{12\sqrt{6}} & -\frac{5}{12\sqrt{6}}&\frac{5}{36\sqrt{2}}&0 \\  
\end{array} \right),\nonumber
\end{align}
\begin{align}
d_{abc}F^a_2F^b_3F^c_6=\left(\begin{array}{ccccc}
\frac{35}{72} &-\frac{5}{8\sqrt{3}}&-\frac{5}{8\sqrt{3}}&\frac{5}{24}& \frac{5}{12\sqrt{2}}    \\                       
-\frac{5}{8\sqrt{3}} &\frac{5}{72} & \frac{5}{24} &-\frac{5}{24\sqrt{3}} &-\frac{5}{12\sqrt{6}} \\  
-\frac{5}{8\sqrt{3}} &\frac{5}{24} & \frac{5}{72} &-\frac{5}{24\sqrt{3}} &-\frac{5}{12\sqrt{6}} \\  
\frac{5}{24} &-\frac{5}{24\sqrt{3}} & -\frac{5}{24\sqrt{3}}&-\frac{5}{72}&\frac{5}{36\sqrt{2}} \\  
\frac{5}{12\sqrt{2}} &-\frac{5}{12\sqrt{6}} & -\frac{5}{12\sqrt{6}}&\frac{5}{36\sqrt{2}}&0 \\  
\end{array} \right),
d_{abc}F^a_2F^b_4F^c_5=\left(\begin{array}{ccccc}
\frac{35}{72} &\frac{5}{8\sqrt{3}}&-\frac{5}{8\sqrt{3}}&-\frac{5}{24}& -\frac{5}{12\sqrt{2}}    \\                       
\frac{5}{8\sqrt{3}} &\frac{5}{72} & -\frac{5}{24} &-\frac{5}{24\sqrt{3}} &-\frac{5}{12\sqrt{6}} \\  
-\frac{5}{8\sqrt{3}} &-\frac{5}{24} & \frac{5}{72} &\frac{5}{24\sqrt{3}} &\frac{5}{12\sqrt{6}} \\  
-\frac{5}{24} &-\frac{5}{24\sqrt{3}} & \frac{5}{24\sqrt{3}}&-\frac{5}{72}&\frac{5}{36\sqrt{2}} \\  
-\frac{5}{12\sqrt{2}} &-\frac{5}{12\sqrt{6}} & \frac{5}{12\sqrt{6}}&\frac{5}{36\sqrt{2}}&0 \\  
\end{array} \right),\nonumber
\end{align}
\begin{align}
d_{abc}F^a_2F^b_4F^c_6=\left(\begin{array}{ccccc}
\frac{35}{72} &\frac{5}{8\sqrt{3}}&\frac{5}{8\sqrt{3}}&\frac{5}{24}& \frac{5}{12\sqrt{2}}      \\                       
\frac{5}{8\sqrt{3}} &\frac{5}{72} & \frac{5}{24} &\frac{5}{24\sqrt{3}} &\frac{5}{12\sqrt{6}} \\  
\frac{5}{8\sqrt{3}} &\frac{5}{24} & \frac{5}{72} &\frac{5}{24\sqrt{3}} &\frac{5}{12\sqrt{6}} \\  
\frac{5}{24} &\frac{5}{24\sqrt{3}} & \frac{5}{24\sqrt{3}}&-\frac{5}{72}&\frac{5}{36\sqrt{2}} \\  
\frac{5}{12\sqrt{2}} &\frac{5}{12\sqrt{6}} & \frac{5}{12\sqrt{6}}&\frac{5}{36\sqrt{2}}&0 \\  
\end{array} \right),
d_{abc}F^a_2F^b_5F^c_6=\left(\begin{array}{ccccc}
-\frac{5}{36}       &    0     & 0           & 0                                 & 0                                       \\  
0                  & -\frac{5}{36}    & 0           &  0                    &      0                                    \\  
0                  &            0          & \frac{25}{36}   &\frac{5}{6\sqrt{3}}  & -\frac{5}{6\sqrt{6}}    \\ 
0                  & 0               & \frac{5}{6\sqrt{3}} & \frac{5}{36}     & -\frac{5}{18\sqrt{2}}        \\
0                  & 0               & -\frac{5}{6\sqrt{6}}   & -\frac{5}{18\sqrt{2}}      &  0                    \\                                                      
\end{array} \right),\nonumber
\end{align}
\begin{align}
d_{abc}F^a_3F^b_4F^c_5=\left(\begin{array}{ccccc}
-\frac{5}{36}       &    0     & 0           & 0                                 & 0                                       \\  
0                  &  \frac{25}{36}    & 0           &  -\frac{5}{6\sqrt{3}}  & \frac{5}{6\sqrt{6}}        \\  
0                  &            0              & -\frac{5}{36}       &      0            & 0                          \\ 
0                  & -\frac{5}{6\sqrt{3}} &  0         & \frac{5}{36}     & -\frac{5}{18\sqrt{2}}        \\
0                  & \frac{5}{6\sqrt{6}}  &     0   & -\frac{5}{18\sqrt{2}}      &  0                       \\                                                      
\end{array} \right),\nonumber
d_{abc}F^a_3F^b_4F^c_6=\left(\begin{array}{ccccc}
-\frac{5}{36}       &    0     & 0           & 0                                 & 0                                       \\  
0                  &  \frac{25}{36}    & 0           &  \frac{5}{6\sqrt{3}}  & -\frac{5}{6\sqrt{6}}        \\  
0                  &            0              & -\frac{5}{36}       &      0            & 0                          \\ 
0                  & \frac{5}{6\sqrt{3}} &  0         & \frac{5}{36}     & -\frac{5}{18\sqrt{2}}        \\
0                  & -\frac{5}{6\sqrt{6}}  &     0   & -\frac{5}{18\sqrt{2}}      &  0                       \\                                                      
\end{array} \right),\nonumber 
\end{align}
\begin{equation}
d_{abc}F^a_3F^b_5F^c_6=\left(\begin{array}{ccccc}
-\frac{5}{36}  &    0             & 0                      & 0                    & 0                              \\  
0                  &   -\frac{5}{36}         & 0              & 0                             & 0                   \\  
0                  &            0              & -\frac{5}{36}    &      0                   & 0                    \\ 
0                  &             0         &  0                       &  \frac{35}{36}    & \frac{5}{9\sqrt{2}} \\
0                  &       0             &     0                 &   \frac{5}{9\sqrt{2}}     &   0               \\                                                      
\end{array} \right),
d_{abc}F^a_4F^b_5F^c_6=\left(\begin{array}{ccccc}
-\frac{5}{36}       &    0     & 0           & 0                                 & 0                                       \\  
0                  &   -\frac{5}{36}            & 0              & 0                             & 0                   \\  
0                  &            0              & -\frac{5}{36}       &      0            & 0                          \\ 
0                  &             0         &  0                        & -\frac{5}{36}             & 0               \\
0                  &       0             &     0                 &       0         &   \frac{10}{9}                  \\                                                      
\end{array} \right). 
\end{equation}
We can prove that $\sum_{i<j<k}^{6}d_{abc}F^a_i F^b_j F^c_k$ is invariant to  SU(3), which means that it commutes with the generators of dibaryon, $F^a$ = 1/2 ( ${\lambda}^a_1$ + ${\lambda}^a_2$ + ${\lambda}^a_3$ + ${\lambda}^a_4$ + ${\lambda}^a_5$ + ${\lambda}^a_6$ ), as it is proportional to the identity operator as can be seen in the following:  
\begin{align}
\sum_{i<j<k}^{6}d_{abc}F^a_i F^b_j F^c_k=\left(\begin{array}{ccccc}
\frac{20}{9}       &    0     & 0           & 0                                 & 0                                       \\  
0                  &   \frac{20}{9}            & 0              & 0                             & 0                   \\  
0                  &            0              & \frac{20}{9}       &      0            & 0                          \\ 
0                  &             0         &  0                        & \frac{20}{9}             & 0               \\
0                  &       0             &     0                 &       0         &   \frac{20}{9}                  \\                                                      
\end{array} \right). 
\end{align}

The matrix of $f_{abc}F^a_i F^b_j F^c_k$ is given by,
\begin{equation}
f_{abc}F^a_1F^b_2F^c_3=\left(\begin{array}{ccccc}
0                                    & \frac{i\sqrt{3}}{4}  & 0           & 0                                 & 0             \\  
-\frac{i\sqrt{3}}{4} & 0                                     & 0              & 0                             & 0                \\  
0                                  &            0                          &         0         & \frac{i\sqrt{3}}{4}    &0       \\ 
0                               &             0                   &-\frac{i\sqrt{3}}{4}       & 0           & 0               \\
0                        &       0             &   0      &       0   &   0                                                                      \\
\end{array} \right),\nonumber 
f_{abc}F^a_1F^b_2F^c_4=\left(\begin{array}{ccccc}
0                                    & -\frac{i\sqrt{3}}{4}  & 0           & 0                                 & 0             \\  
\frac{i\sqrt{3}}{4} & 0                                     & 0              & 0                             & 0                \\  
0                      &            0        &  0                &  \frac{i}{4\sqrt{3}}   & -\frac{i}{\sqrt{6}}         \\ 
0                  &             0       &-\frac{i}{4\sqrt{3}}       & 0           & 0                                     \\
0              &       0             &   \frac{i}{\sqrt{6}}       &       0   &   0                                       \\
\end{array} \right), 
\end{equation}
\begin{equation}
f_{abc}F^a_1F^b_2F^c_5=\left(\begin{array}{ccccc} 
0                        & 0         & 0           & \frac{i}{4}         & \frac{i}{2\sqrt{2}}                     \\  
0                  &    0                 & -\frac{i}{4}           & 0                             & 0                   \\  
0               &  \frac{i}{4}          &  0           & -\frac{i}{2\sqrt{3}}    & \frac{i}{2\sqrt{6}}          \\ 
-\frac{i}{4}      &         0                   &\frac{i}{2\sqrt{3}}                    & 0           & 0               \\
-\frac{i}{2\sqrt{2}}     &       0             &   -\frac{i}{2\sqrt{6}}        &       0   &   0                  \\                                                                      
\end{array} \right),\nonumber 
f_{abc}F^a_1F^b_2F^c_6=\left(\begin{array}{ccccc}
0                        & 0         & 0           & -\frac{i}{4}         & \frac{i}{2\sqrt{2}}                     \\  
0                  &    0                 & \frac{i}{4}           & 0                             & 0                   \\  
0               &  -\frac{i}{4}          &  0           & -\frac{i}{2\sqrt{3}}    & \frac{i}{2\sqrt{6}}          \\ 
\frac{i}{4}      &         0                   &\frac{i}{2\sqrt{3}}                    & 0           & 0               \\
\frac{i}{2\sqrt{2}}     &       0             &   -\frac{i}{2\sqrt{6}}        &       0   &   0                  \\                                                                      
\end{array} \right), 
\end{equation}
\begin{equation}
f_{abc}F^a_1F^b_3F^c_4=\left(\begin{array}{ccccc}
0                                    & \frac{i\sqrt{3}}{4}  & 0           & 0                                 & 0             \\  
-\frac{i\sqrt{3}}{4} & 0                                     & 0              & 0                             & 0                \\  
0                      &            0        &  0                &  -\frac{i}{4\sqrt{3}}   & -\frac{i}{2\sqrt{6}}         \\ 
0           &             0       & \frac{i}{4\sqrt{3}}       & 0          &\frac{i}{2\sqrt{2}}                        \\                                     
0              &       0             &   \frac{i}{2\sqrt{6}}   & -\frac{i}{2\sqrt{2}}   &   0                             \\                                       
\end{array} \right), 
f_{abc}F^a_1F^b_3F^c_5=\left(\begin{array}{ccccc} 
0                        & 0         & 0           & -\frac{i}{4}         & \frac{i}{4\sqrt{2}}                     \\  
0                &    0          & \frac{i}{4}         & 0        & -\frac{i\sqrt{3}}{4\sqrt{2}}                 \\  
0               &  -\frac{i}{4}          &  0           & \frac{i}{2\sqrt{3}}    & \frac{i}{4\sqrt{6}}          \\ 
\frac{i}{4}      &         0          &-\frac{i}{2\sqrt{3}}          & 0           & -\frac{i}{4\sqrt{2}}          \\
-\frac{i}{4\sqrt{2}} &\frac{i\sqrt{3}}{4\sqrt{2}} &-\frac{i}{4\sqrt{6}} & \frac{i}{4\sqrt{2}} &  0   \\                                                                      
\end{array} \right),\nonumber 
\end{equation}
\begin{equation}
f_{abc}F^a_1F^b_3F^c_6=\left(\begin{array}{ccccc}
0                        & 0         & 0           & -\frac{i}{4}         & -\frac{i}{4\sqrt{2}}                     \\  
0                &    0          & -\frac{i}{4}         & 0        & \frac{i\sqrt{3}}{4\sqrt{2}}                 \\  
0               &  \frac{i}{4}          &  0           & \frac{i}{2\sqrt{3}}    & \frac{i}{4\sqrt{6}}          \\ 
-\frac{i}{4}      &         0          &-\frac{i}{2\sqrt{3}}          & 0           & -\frac{i}{4\sqrt{2}}          \\
\frac{i}{4\sqrt{2}} &-\frac{i\sqrt{3}}{4\sqrt{2}} &-\frac{i}{4\sqrt{6}} & \frac{i}{4\sqrt{2}} &  0   \\                                                                      
\end{array} \right),\nonumber 
f_{abc}F^a_1F^b_4F^c_5=\left(\begin{array}{ccccc} 
0                        & 0         & 0           & \frac{i}{4}         & -\frac{i}{4\sqrt{2}}                     \\  
0                &    0          & \frac{i}{4}  & \frac{i}{2\sqrt{3}}      & \frac{i}{4\sqrt{6}}         \\  
0               &  -\frac{i}{4}          &  0           & 0            & -\frac{i\sqrt{3}}{4\sqrt{2}}          \\ 
-\frac{i}{4}  & -\frac{i}{2\sqrt{3}}  &  0          & 0               &      \frac{i}{4\sqrt{2}}          \\
\frac{i}{4\sqrt{2}} &-\frac{i}{4\sqrt{6}} &\frac{i\sqrt{3}}{4\sqrt{2}} & -\frac{i}{4\sqrt{2}} &  0   \\                                                                      
\end{array} \right),\nonumber 
\end{equation}
\begin{equation}
f_{abc}F^a_1F^b_4F^c_6=\left(\begin{array}{ccccc} 
0                        & 0         & 0           & -\frac{i}{4}         & \frac{i}{4\sqrt{2}}                     \\  
0                &    0          & -\frac{i}{4}  & -\frac{i}{2\sqrt{3}}      & -\frac{i}{4\sqrt{6}}         \\  
0               &  \frac{i}{4}          &  0           & 0            & -\frac{i\sqrt{3}}{4\sqrt{2}}          \\ 
\frac{i}{4}  & \frac{i}{2\sqrt{3}}  &  0          & 0               &      \frac{i}{4\sqrt{2}}          \\
-\frac{i}{4\sqrt{2}} &\frac{i}{4\sqrt{6}} &\frac{i\sqrt{3}}{4\sqrt{2}} & -\frac{i}{4\sqrt{2}} &  0   \\                                                                      
\end{array} \right),\nonumber 
f_{abc}F^a_1F^b_5F^c_6=\left(\begin{array}{ccccc}
0                        & 0         & 0           & \frac{i}{4}         & \frac{i}{2\sqrt{2}}                     \\  
0                &    0          & \frac{i}{4}   & \frac{i}{2\sqrt{3}} & -\frac{i}{2\sqrt{6}}              \\  
0               &  -\frac{i}{4}          &  0           & 0                &    0                                 \\ 
-\frac{i}{4}      &   -\frac{i}{2\sqrt{3}}   &    0          & 0           &  0          \\
-\frac{i}{2\sqrt{2}} &\frac{i}{2\sqrt{6}} &  0 &  0 &  0   \\                                                                      
\end{array} \right),
\end{equation}
\begin{equation}
f_{abc}F^a_2F^b_3F^c_4=\left(\begin{array}{ccccc}
0                                    & \frac{i\sqrt{3}}{4}  & 0           & 0                                 & 0             \\  
-\frac{i\sqrt{3}}{4} & 0                                     & 0              & 0                             & 0                \\  
0                      &            0        &  0                &  -\frac{i}{4\sqrt{3}}   & -\frac{i}{2\sqrt{6}}         \\ 
0           &             0       & \frac{i}{4\sqrt{3}}       & 0          &-\frac{i}{2\sqrt{2}}                        \\                                     
0              &       0             &   \frac{i}{2\sqrt{6}}   & \frac{i}{2\sqrt{2}}   &   0                             \\                                       
\end{array} \right), 
f_{abc}F^a_2F^b_3F^c_5=\left(\begin{array}{ccccc} 
0                        & 0         & 0           & -\frac{i}{4}         & \frac{i}{4\sqrt{2}}                     \\  
0                &    0          & \frac{i}{4}         & 0        & \frac{i\sqrt{3}}{4\sqrt{2}}                 \\  
0               &  -\frac{i}{4}          &  0           & \frac{i}{2\sqrt{3}}    & \frac{i}{4\sqrt{6}}          \\ 
\frac{i}{4}      &         0          &-\frac{i}{2\sqrt{3}}          & 0           & \frac{i}{4\sqrt{2}}          \\
-\frac{i}{4\sqrt{2}} &-\frac{i\sqrt{3}}{4\sqrt{2}} &-\frac{i}{4\sqrt{6}} & -\frac{i}{4\sqrt{2}} &  0   \\                                                                      
\end{array} \right),\nonumber 
\end{equation}
\begin{equation}
f_{abc}F^a_2F^b_3F^c_6=\left(\begin{array}{ccccc}
0                        & 0         & 0           & \frac{i}{4}         & -\frac{i}{4\sqrt{2}}                     \\  
0                &    0          & -\frac{i}{4}         & 0        & -\frac{i\sqrt{3}}{4\sqrt{2}}                 \\  
0               &  \frac{i}{4}          &  0           & \frac{i}{2\sqrt{3}}    & \frac{i}{4\sqrt{6}}          \\ 
-\frac{i}{4}      &         0          &-\frac{i}{2\sqrt{3}}          & 0           & \frac{i}{4\sqrt{2}}          \\
\frac{i}{4\sqrt{2}} &\frac{i\sqrt{3}}{4\sqrt{2}} &-\frac{i}{4\sqrt{6}} & -\frac{i}{4\sqrt{2}} &  0   \\                                                                      
\end{array} \right),\nonumber 
f_{abc}F^a_2F^b_4F^c_5=\left(\begin{array}{ccccc} 
0                        & 0         & 0           & \frac{i}{4}         & -\frac{i}{4\sqrt{2}}                     \\  
0                &    0          & \frac{i}{4}  & -\frac{i}{2\sqrt{3}}      & -\frac{i}{4\sqrt{6}}         \\  
0               &  -\frac{i}{4}          &  0           & 0            & -\frac{i\sqrt{3}}{4\sqrt{2}}          \\ 
-\frac{i}{4}  & \frac{i}{2\sqrt{3}}  &  0          & 0               &      -\frac{i}{4\sqrt{2}}          \\
\frac{i}{4\sqrt{2}} &\frac{i}{4\sqrt{6}} &\frac{i\sqrt{3}}{4\sqrt{2}} &\frac{i}{4\sqrt{2}} &  0   \\                                                                      
\end{array} \right),\nonumber 
\end{equation}
\begin{equation}
f_{abc}F^a_2F^b_4F^c_6=\left(\begin{array}{ccccc} 
0                        & 0         & 0           & -\frac{i}{4}         & \frac{i}{4\sqrt{2}}                     \\  
0                &    0          & -\frac{i}{4}  & \frac{i}{2\sqrt{3}}      & \frac{i}{4\sqrt{6}}         \\  
0               &  \frac{i}{4}          &  0           & 0            & -\frac{i\sqrt{3}}{4\sqrt{2}}          \\ 
\frac{i}{4}  & -\frac{i}{2\sqrt{3}}  &  0          & 0               &      -\frac{i}{4\sqrt{2}}          \\
-\frac{i}{4\sqrt{2}} &-\frac{i}{4\sqrt{6}} &\frac{i\sqrt{3}}{4\sqrt{2}} & \frac{i}{4\sqrt{2}} &  0   \\                                                                      
\end{array} \right),\nonumber 
f_{abc}F^a_2F^b_5F^c_6=\left(\begin{array}{ccccc}
0                        & 0         & 0           & \frac{i}{4}         & \frac{i}{2\sqrt{2}}                     \\  
0                &    0          & \frac{i}{4}   & -\frac{i}{2\sqrt{3}} & \frac{i}{2\sqrt{6}}              \\  
0               &  -\frac{i}{4}          &  0           & 0                &    0                                 \\ 
-\frac{i}{4}      &   \frac{i}{2\sqrt{3}}   &    0          & 0           &  0          \\
-\frac{i}{2\sqrt{2}} &-\frac{i}{2\sqrt{6}} &  0 &  0 &  0   \\                                                                      
\end{array} \right),
\end{equation}
\begin{equation}
f_{abc}F^a_3F^b_4F^c_5=\left(\begin{array}{ccccc} 
0                    & 0         &  -\frac{i\sqrt{3}}{4}        & 0         & 0                       \\  
0                &    0          &  0  & \frac{i}{4\sqrt{3}}      & \frac{i}{2\sqrt{6}}         \\  
\frac{i\sqrt{3}}{4}        &  0         &  0                & 0                   &    0          \\ 
0    & -\frac{i}{4\sqrt{3}}  &  0          & 0               &      \frac{i}{2\sqrt{2}}          \\
0    & -\frac{i}{2\sqrt{6}}  &  0          &   -\frac{i}{2\sqrt{2}}  &      0          \\
\end{array} \right),\nonumber 
f_{abc}F^a_3F^b_4F^c_6=\left(\begin{array}{ccccc}
0                    & 0         &  \frac{i\sqrt{3}}{4}        & 0         & 0                       \\  
0                &    0          &  0  & -\frac{i}{4\sqrt{3}}      & -\frac{i}{2\sqrt{6}}         \\  
-\frac{i\sqrt{3}}{4}        &  0         &  0                & 0                   &    0          \\ 
0    & \frac{i}{4\sqrt{3}}  &  0          & 0               &      \frac{i}{2\sqrt{2}}          \\
0    & \frac{i}{2\sqrt{6}}  &  0          &   -\frac{i}{2\sqrt{2}}  &      0          \\
\end{array} \right),
\end{equation}
\begin{equation}
f_{abc}F^a_3F^b_5F^c_6=\left(\begin{array}{ccccc} 
0                    & 0         &  -\frac{i\sqrt{3}}{4}        & 0         & 0                       \\  
0                &    0          &  0  & \frac{i}{4\sqrt{3}}      & -\frac{i}{\sqrt{6}}         \\  
\frac{i\sqrt{3}}{4}        &  0         &  0                & 0                   &    0          \\ 
0    & -\frac{i}{4\sqrt{3}}  &  0          & 0               &      0          \\
0    & \frac{i}{\sqrt{6}}  &  0          &   0            &      0          \\
\end{array} \right), 
f_{abc}F^a_4F^b_5F^c_6=\left(\begin{array}{ccccc}
0                    & 0         &  \frac{i\sqrt{3}}{4}        & 0         & 0                       \\  
0                &    0          &  0          &  \frac{i\sqrt{3}}{4}             &  0        \\  
-\frac{i\sqrt{3}}{4}        &  0         &  0                & 0                   &    0          \\ 
0    &  -\frac{i\sqrt{3}}{4}  &  0          & 0               &      0          \\
0    & 0         &  0          &   0   &      0          \\
\end{array} \right).
\end{equation}
In this case, it turns out that the SU(3) invariant operator is $f_{abc}F^a_1 F^b_2 F^c_3$ + $f_{abc}F^a_1 F^b_2 F^c_4$ + $f_{abc}F^a_1 F^b_2 F^c_5$ +$f_{abc}F^a_1 F^b_2 F^c_6$ + $f_{abc}F^a_1 F^b_3 F^c_4$ + $f_{abc}F^a_1 F^b_3 F^c_5$ + $f_{abc}F^a_1 F^b_3 F^c_6$ + $f_{abc}F^a_1 F^b_4 F^c_5$ + $f_{abc}F^a_1 F^b_4 F^c_6$ +$f_{abc}F^a_1 F^b_5 F^c_6$ - $f_{abc}F^a_2 F^b_3 F^c_4$ - $f_{abc}F^a_2 F^b_3 F^c_5$ - $f_{abc}F^a_2 F^b_3 F^c_6$ - $f_{abc}F^a_2 F^b_4 F^c_5$ - $f_{abc}F^a_2 F^b_4 F^c_6$ - $f_{abc}F^a_2 F^b_5 F^c_6$ - $f_{abc}F^a_3 F^b_4 F^c_5$ - $f_{abc}F^a_3 F^b_4 F^c_6$ - $f_{abc}F^a_3 F^b_5 F^c_6$ - $f_{abc}F^a_4 F^b_5 F^c_6$, due to the fact that this operator is
\begin{align}
\left(\begin{array}{ccccc}
0       &    0     & 0           & 0                                 & 0                                       \\  
0                  &  0            & 0              & 0                             & 0                   \\  
0                  &            0              &  0       &      0            & 0                          \\ 
0                  &             0         &  0                        &  0             & 0               \\
0                  &       0             &     0                 &       0         &   0                  \\                                                      
\end{array} \right). 
\end{align}

\end{widetext}


\begin{thebibliography}{50}




\bibitem{Bashkanov:2008ih} 
  M.~Bashkanov, C.~Bargholtz, M.~Berlowski, D.~Bogoslawsky, H.~Calen, H.~Clement, L.~Demiroers and E.~Doroshkevich {\it et al.},
  Phys.\ Rev.\ Lett.\  {\bf 102}, 052301 (2009)



\bibitem{Adlarson:2011bh} 
  P.~Adlarson {\it et al.}  [WASA-at-COSY Collaboration],
  Phys.\ Rev.\ Lett.\  {\bf 106}, 242302 (2011)


\bibitem{Adlarson:2012fe} 
  P.~Adlarson {\it et al.}  [WASA-at-COSY Collaboration],
  Phys.\ Lett.\ B {\bf 721}, 229 (2013)


\bibitem{Adlarson:2014pxj} 
  P.~Adlarson {\it et al.}  [WASA-at-COSY Collaboration],
  Phys.\ Rev.\ Lett.\  {\bf 112}, no. 20, 202301 (2014)



\bibitem{Adlarson:2014ozl} 
  P.~Adlarson {\it et al.}  [WASA-at-COSY Collaboration],
  Phys.\ Rev.\ C {\bf 90}, no. 3, 035204 (2014)



\bibitem{Adlarson:2014tcn} 
  P.~Adlarson {\it et al.}  [WASA-at-COSY Collaboration],
  Phys.\ Lett.\ B {\bf 743}, 325 (2015)


\bibitem{Jaffe:1976yi} 
  R.~L.~Jaffe,
  Phys.\ Rev.\ Lett.\  {\bf 38}, 195 (1977)
  [Phys.\ Rev.\ Lett.\  {\bf 38}, 617 (1977)].


\bibitem{Goldman:1989zj} 
  J.~T.~Goldman, K.~Maltman, G.~J.~Stephenson, Jr., K.~E.~Schmidt and F.~Wang,
  Phys.\ Rev.\ C {\bf 39}, 1889 (1989).



\bibitem{Oka:1980ax} 
  M.~Oka and K.~Yazaki,
  Phys.\ Lett.\ B {\bf 90}, 41 (1980).



\bibitem{SilvestreBrac:1992yg} 
  B.~Silvestre- Brac and J.~Leandri,
  Phys.\ Rev.\ D {\bf 45}, 4221 (1992).


\bibitem{Mackenzie:1985vv} 
  P.~B.~Mackenzie and H.~B.~Thacker,
  Phys.\ Rev.\ Lett.\  {\bf 55}, 2539 (1985).


\bibitem{Aerts:1984vv} 
  A.~T.~M.~Aerts and J.~Rafelski,
  Phys.\ Lett.\ B {\bf 148}, 337 (1984).


\bibitem{Balachandran:1983dj} 
  A.~P.~Balachandran, A.~Barducci, F.~Lizzi, V.~G.~J.~Rodgers and A.~Stern,
  Phys.\ Rev.\ Lett.\  {\bf 52}, 887 (1984).



\bibitem{Straub:1988mz} 
  U.~Straub, Z.~Y.~Zhang, K.~Brauer, A.~Faessler and S.~B.~Khadkikar,
  Phys.\ Lett.\ B {\bf 200}, 241 (1988).



\bibitem{Stancu:1998ca} 
  F.~Stancu, S.~Pepin and L.~Y.~Glozman,
  Phys.\ Rev.\ D {\bf 57}, 4393 (1998)

\bibitem{Inoue:2010es} 
  T.~Inoue {\it et al.}  [HAL QCD Collaboration],
  Phys.\ Rev.\ Lett.\  {\bf 106}, 162002 (2011)
  [arXiv:1012.5928 [hep-lat]].

\bibitem{Beane:2010hg} 
  S.~R.~Beane {\it et al.}  [NPLQCD Collaboration],
  Phys.\ Rev.\ Lett.\  {\bf 106}, 162001 (2011)
  [arXiv:1012.3812 [hep-lat]].

\bibitem{Maltman:1986xz} 
  K.~Maltman,
  Nucl.\ Phys.\ A {\bf 459}, 475 (1986).


\bibitem{Goldman:1987ma} 
  J.~T.~Goldman, K.~Maltman, G.~J.~Stephenson, Jr., K.~E.~Schmidt and F.~Wang,
  Phys.\ Rev.\ Lett.\  {\bf 59}, 627 (1987).


\bibitem{Pepin:1998ih} 
  S.~Pepin and F.~Stancu,
  Phys.\ Rev.\ D {\bf 57}, 4475 (1998)
  [hep-ph/9710528].


\bibitem{Zouzou:1986qh} 
  S.~Zouzou, B.~Silvestre-Brac, C.~Gignoux and J.~M.~Richard,
  Z.\ Phys.\ C {\bf 30}, 457 (1986).

\bibitem{SilvestreBrac:1993ry} 
  B.~Silvestre-Brac and C.~Semay,
  Z.\ Phys.\ C {\bf 59}, 457 (1993).

\bibitem{Brink:1998as} 
  D.~M.~Brink and F.~Stancu,
  Phys.\ Rev.\ D {\bf 57}, 6778 (1998).

\bibitem{Park:2013fda} 
  W.~Park and S.~H.~Lee,
  Nucl.\ Phys.\ A {\bf 925}, 161 (2014)


\bibitem{Leandri:1995zm} 
  J.~Leandri and B.~Silvestre-Brac,
  Phys.\ Rev.\ D {\bf 51}, 3628 (1995).


\bibitem{Bhaduri:1981pn} 
  R.~K.~Bhaduri, L.~E.~Cohler and Y.~Nogami,
  Nuovo Cim.\ A {\bf 65}, 376 (1981).

\bibitem{Pepin:2001is} 
  S.~Pepin and F.~Stancu,
  Phys.\ Rev.\ D {\bf 65}, 054032 (2002)

\bibitem{Dmitrasinovic:2001nu} 
  V.~Dmitrasinovic,
  Phys.\ Lett.\ B {\bf 499}, 135 (2001)

\bibitem{Stancu:1991rc} 
  F.~Stancu,
  Oxford Stud.\ Nucl.\ Phys.\  {\bf 19}, 1 (1996).



\bibitem{Stancu:1999qr} 
  F.~Stancu and S.~Pepin,
  Few Body Syst.\  {\bf 26}, 113 (1999).





\bibitem{Aerts:1977rw} 
  A.~T.~M.~Aerts, P.~J.~G.~Mulders and J.~J.~de Swart,
  Phys.\ Rev.\ D {\bf 17}, 260 (1978).


\bibitem{Dmitrasinovic:2001nv} 
  V.~Dmitrasinovic,
  J.\ Math.\ Phys.\  {\bf 42}, 991 (2001)
  [J.\ Math.\ Phys.\  {\bf 45}, 2988 (2004)]
  [hep-ph/0101008].



\end{thebibliography}
\end{document}